\documentclass[twocolumn]{aastex701}
\usepackage{amsmath}
\usepackage{multirow}

\graphicspath{{./}{figures/}}
\usepackage{color,subfigure,lineno,booktabs} 

\definecolor{firebrick}{rgb}{0.7, 0.13, 0.13}

\newcommand{\OIIIab}{[O{\sc iii}]\,$\lambda\lambda$4959,5007}
\newcommand{\Hb}{H$\beta$}
\newcommand{\Ha}{H$\alpha$}

\def\CIV{C\,{\sc iv}}

\def\CIV{C\,{\sc iv}}
\def\MgII{Mg\,{\sc ii}}

\def\FeII{Fe\,{\sc ii}}
\def\OIII{[O\,{\sc iii}]}

\shorttitle{NEXUS: MSA LRDs} %
\shortauthors{Pan et~al.}

\begin{document}

\title{NEXUS: Abundance, Environments, and Spectral Diversity of Little Red Dots from the NIRSpec MSA Sample}

\correspondingauthor{Zhiwei Pan}
\email{zhiweip@illinois.edu}

\author[0000-0003-0230-6436]{Zhiwei Pan}
\affiliation{Department of Astronomy, University of Illinois Urbana-Champaign, Urbana, IL 61801, USA}
\email{zhiweip@illinois.edu}

\author[0000-0001-5105-2837]{Ming-Yang Zhuang}
\affiliation{Department of Astronomy, University of Illinois Urbana-Champaign, Urbana, IL 61801, USA}
\email[show]{mingyang@illinois.edu}

\author[0000-0003-1659-7035]{Yue Shen}
\affiliation{Department of Astronomy, University of Illinois Urbana-Champaign, Urbana, IL 61801, USA}
\affiliation{National Center for Supercomputing Applications, University of Illinois Urbana-Champaign, Urbana, IL 61801, USA}
\email[show]{shenyue@illinois.edu}

\author[0000-0002-7633-431X]{Feige Wang} \email{fgwang@umich.edu}
\affiliation{Department of Astronomy, University of Michigan, 1085 S University, Ann Arbor, MI 48109, USA}

\author[0000-0002-5612-3427]{Jenny E. Greene}
\email{jgreene@astro.princeton.edu}
\affiliation{Department of Astrophysical Sciences, Princeton University, 4 Ivy Lane, Princeton, NJ 08544, USA}

\author[0000-0002-6523-9536]{Adam J.~Burgasser}
\affiliation{UC San Diego, La Jolla, CA 92093, USA}
\email{aburgasser@ucsd.edu}

\author[0000-0002-1605-915X]{Junyao Li}
\affiliation{Department of Astronomy, University of Illinois Urbana-Champaign, Urbana, IL 61801, USA}
\email{junyaoli@illinois.edu}

\author[0000-0002-8501-3518]{Zachary Stone}
\affiliation{Department of Astronomy, University of Illinois Urbana-Champaign, Urbana, IL 61801, USA}
\email{stone28@illinois.edu}

\author[0000-0001-8638-2780]{Padmavathi Venkatraman}
\affiliation{Department of Astronomy, University of Illinois Urbana-Champaign, Urbana, IL 61801, USA}
\email{pv10@illinois.edu}


\begin{abstract}
We present a comprehensive study of Little Red Dots (LRDs) at $2.3<z<7.4$ using NIRCam photometry and NIRSpec MSA/PRISM spectra from the ongoing NEXUS program. Photometric selection combining several commonly adopted methods yields a high completeness ($\sim 85\%$) of LRD selection over this redshift range and for a flux limit of F444W$<26$. The overall purity is $\sim 60\%$, with contamination from emission-line galaxies and normal active galactic nuclei (AGNs), as well as dwarf stars. Most ($\gtrsim90\%$) of the spectroscopically confirmed LRDs have robust broad-line detection. Our spectroscopic sample of 36 LRDs displays the full range of spectral diversity of LRDs. It includes objects with extreme Balmer breaks similar to the LRD “Cliff”, as well as objects with moderately reddened rest-optical continua that can be fit with low-temperature blackbody components in the recent BH* model framework. The broad H$\alpha$ emission is correlated with the continuum 5100\AA\ emission, suggesting common origins for these emission components; the narrow \OIII{} emission, however, is poorly correlated with the optical continuum. We do not find evidence of redshift evolution of these spectral properties. The space density of LRDs declines toward $z\approx2$, opposite to the trend for normal AGNs, although low-luminosity LRDs at $z\sim2$--4 may be more abundant than currently probed by ground-based searches. The clustering of LRDs suggest that they live in dark matter halos of several $\times 10^{11}\,h^{-1}M_\odot$ albeit with large uncertainties. Overall these results are consistent with recent observations of LRDs and the emerging picture of accreting SMBHs enshrouded in dense gas envelopes as the origin of LRDs.  
\end{abstract}
\keywords{Active galactic nuclei ---  Galaxy evolution --- High-redshift galaxies --- Quasars --- Surveys}

\section{Introduction}\label{sec:introduction}

The James Webb Space Telescope (JWST) has revealed a remarkable variety of active galactic nuclei (AGNs), reshaping our view of the early Universe \citep[e.g.,][]{Harikane_2023,YangJinyi_2023,Maiolino_2024,Furtak_2024,JiangDanyang_2025,FuShuqi_2025,FeiQinyue_2026}. Among its particularly intriguing discoveries is a population of compact, red sources with unusual spectral energy distributions (SEDs). These objects exhibit a characteristic ``V-shape'' SED, with blue continua in the rest-frame ultraviolet (UV) and markedly red colors in the rest-frame optical. Their compact morphologies and red rest-optical colors have led to the name ``Little Red Dots'' \citep[LRDs; e.g.,][]{Labbe_2023,Furtak_2023,Barro_2024,Matthee_2024}. LRDs appear to be common over a wide redshift range, from $z\sim4$ to $z\sim9$ \citep[e.g.,][]{Greene_2024_Vshape_UNCOVER,Barro_2024,Kokorev_2024_Vshape,Kocevski_2025_Slope}.

Spectroscopic follow-up of photometrically selected LRDs has further revealed that $\gtrsim 70\%$ show broad Balmer emission lines \citep[e.g.,][]{Harikane_2023,Greene_2024_Vshape_UNCOVER,Killi_2024,Matthee_2024,WangBingjie_2025,Hviding_2025,ZhuangMingyang_2026_NEXUS_LRD}. However, LRDs differ from typical broad-line AGNs in several important ways. They generally lack strong broad high-ionization lines in the rest-frame UV, are weak or undetected in X-rays \citep[e.g.,][]{Ananna_2024,Kokubo_2025,Maiolino_2025,YueMinghao_2024}, show no detectable radio emission \citep[e.g.,][]{Latif_2025,Gloudemans_2025,Mazzolari_2026}, and are weak or undetected in the far-infrared, implying little hot-dust emission \citep[e.g.,][]{Williams_2024,Setton_2025,Leung_2025,XiaoMengyuan_2025}. In addition, multi-epoch observations have found little or no variability in many LRDs \citep[e.g.,][]{ZhangZijian_2025,Kokubo_2025,Stone_2025,Tee_2025,LiuZhaoran_2026}.

Despite these rapid observational advances, the physical nature of LRDs remains uncertain. One interpretation is that LRDs are dominated by stellar emission from very compact, massive galaxies. In this picture, their red rest-frame optical colors arise from strong Balmer breaks and/or dusty star formation, implying stellar masses of $\gtrsim 10^{10}~M_\odot$ in systems with very small effective radii and extraordinarily high stellar mass densities \citep[e.g.,][]{Labbe_2023,Baggen_2023,Guia_2024}. JWST/MIRI observations of several LRDs have revealed relatively flat rest-frame mid-infrared SEDs, in some cases suggestive of a $1.6~\mu$m stellar bump rather than hot-dust emission from an AGN. This has been interpreted as additional support for a stellar-dominated origin \citep[e.g.,][]{Perez-Gonzalez_2024,Williams_2024,WangBingjie_2025}. This scenario is also consistent with the general X-ray weakness of the population and with the frequent detection of Balmer breaks in LRD spectra \citep[e.g.,][]{Kokorev_2024,WangBingjie_2024_RUBIES,WangBingjie_2025}.

An alternative explanation is that LRDs are powered by reddened accreting SMBHs. In this framework, their compact morphologies and broad Balmer lines are naturally explained. For example, \citet{Greene_2024_Vshape_UNCOVER} modeled the SEDs of LRDs with a scattered AGN component dominating the rest-frame UV and a heavily reddened AGN component contributing in the rest-frame optical. \citet{LiZhengrong_2025} showed that the flat mid-infrared continua of LRDs may arise naturally from extended dust and gas distributions in the AGN torus, although it remains unclear whether this explanation applies universally. Meanwhile, the observed X-ray weakness of LRDs does not necessarily argue against AGN, as it may reflect super-Eddington accretion and associated radiative effects \citep[e.g.,][]{Madau_2024,Inayoshi_2025}.

Recent work has shown that both dusty-galaxy and conventional AGN models are likely incomplete, as they have difficulty explaining the Balmer-break strengths of the most extreme LRDs discovered to date, such as \textit{MoM-BH*-1} \citep{Naidu_2025} and the Cliff \citep{deGraaff_2025_Cliff}. A new class of models has therefore been proposed in which extremely dense gas enshrouds the SMBH \citep{Inayoshi&Maiolino_2025}. Such models can account for the extreme Balmer breaks and Balmer absorption features frequently seen in LRDs, but rarely observed in normal low-redshift AGNs \citep{Inayoshi_2025,JiXihan_2025,ZhuangMingyang_2026_NEXUS_LRD}. 

This scenario is further developed in the BH* model \citep{Naidu_2025,deGraaff_2025_Cliff}, in which turbulent gas forms a dust-free, stellar-atmosphere-like cocoon around an accreting SMBH. Accretion onto the black hole provides the primary energy source, while the surrounding gaseous cocoon reprocesses the emergent radiation into a blackbody-like optical SED \citep{deGraaff_2025,Sun_2026_BH*}. In addition, radiative-transfer effects, such as electron scattering and resonant scattering in a dense medium, may contribute to the observed broadening of the Balmer lines \citep{Naidu_2025,Rusakov_2026,Chang_2026}. Such dense gas may also explain the \FeII{} emission observed in some LRDs \citep{Tripodi_2025,Torralba_2026,Sun_2026_BH*}. Recent variability measurements of two gravitationally lensed LRDs further support the presence of distinct SMBHs embedded within stellar-atmosphere-like envelopes \citep{ZhangZijian_2025_LRDvariabilityBH*}.

Regardless of their physical origin, deep multi-band photometry provides an efficient route to identifying large samples of LRD candidates. Existing JWST searches at $z>3$ typically exploit the distinctive combination of compact morphology and red near-infrared colors, or equivalently the ``V-shape'' SED produced by blue rest-UV continua together with red rest-optical emission \citep[e.g.,][]{Kokorev_2024_Vshape,Akins_2025_Cosmos_Red}. When reliable photometric redshifts are available, the selection can also be performed using continuum slopes measured separately in the rest-UV and rest-optical regimes \citep{Kocevski_2025_Slope}. 

However, different photometric methods are designed to capture different aspects of the LRD population, and their completeness and purity can vary strongly with redshift, survey depth, filter coverage, and intrinsic SED shape. A spectroscopically confirmed LRD sample is therefore essential for quantifying selection biases, identifying contaminants, and recovering LRDs missed by broadband color cuts. For example, \citet{Hviding_2025} compared their spectroscopically identified LRD sample in the RUBIES survey with published photometric searches and found that existing photometric selections have high accuracy but recover only $\sim35$--$62\%$ of the RUBIES spectroscopic LRDs at $z>3.1$.

A well-characterized spectroscopic sample is critical for measuring the LRD luminosity function and space density, and for constructing a purer sample to investigate their physical nature. The NEXUS survey provides a unique opportunity to build such a sample. As a multi-cycle JWST Treasury program, NEXUS is obtaining cadenced NIRCam imaging and WFSS spectroscopy over a contiguous $\sim0.1~{\rm deg}^2$ field around the North Ecliptic Pole, referred to as NEXUS-Wide, in three yearly epochs during 2024--2028 \citep{ShenYue_2024_NEXUS}. Its central $\sim50~{\rm arcmin}^2$ region, NEXUS-Deep, will be revisited with NIRCam imaging and NIRSpec/MSA spectroscopy for 18 epochs at a cadence of approximately two months \citep{ZhuangMingyang_2026_NEXUS_QR}. In this work, we use the NEXUS data to construct a photometric LRD sample by combining several commonly used selection methods. We then identify a spectroscopic LRD sample to quantify the completeness and purity of these selections. We aim to characterize the spectral and physical properties of the confirmed LRDs and interpret their diversity in the context of the BH* scenario.

The layout of this paper is as follows. In Section~\ref{sec:data}, we describe the NEXUS data, the photometric selection methods, the spectroscopic identification of LRDs, and the resulting completeness and purity of the photometric selections. In Section~\ref{sec:results}, we present the properties of the LRDs, including their spectral properties, the physical properties of the accreting SMBHs and their potential underlying host galaxies, their abundance, and their clustering properties. In Section~\ref{sec:discuss}, we present composite LRD spectra for different SED subclasses and interpret their spectral diversity within the recent BH* model framework. We summarize our results in Section~\ref{sec:sum}. Throughout this paper, magnitudes are given in the AB system. We adopt a $\Lambda$-dominated flat cosmology with $H_0=70~{\rm km~s^{-1}~Mpc^{-1}}$, $\Omega_m=0.3$, and $\Omega_\Lambda=0.7$.


\section{Data and Samples}{} \label{sec:data}

\subsection{NEXUS Data}\label{sec:NEXUS}

NEXUS is a JWST Multi-Cycle (Cycles 3--5) GO Treasury imaging and spectroscopic survey around the North Ecliptic Pole during 2024--2028. It consists of two overlapping tiers in area and depth: a Wide tier covering $\sim0.1\,{\rm deg^2}$ ($\sim400\,{\rm arcmin}^2$) and a Deep tier covering the central $\sim50\,{\rm arcmin}^2$ \citep{ShenYue_2024_NEXUS}. The Wide tier obtains NIRCam/WFSS 2.4--5\,\micron\ grism spectroscopy over three annual epochs, together with NIRCam imaging in the broad bands F090W, F115W, F150W, F200W, F356W, and F444W. The Deep tier carries out high-multiplexing NIRSpec/MSA PRISM spectroscopy over multiple epochs, accompanied by repeated NIRCam imaging. In both tiers, additional parallel imaging with MIRI and extra NIRCam filters is also obtained, although the MIRI observations cover only a fraction of the primary NEXUS footprint because they are carried out in coordinated parallel mode \citep{ShenYue_2024_NEXUS}.

In this work, we use the NIRCam imaging data from Wide epoch 1, Deep epochs 1--5, and Wide epoch 2, including the six broad bands F090W, F115W, F150W, F200W, F356W, and F444W, as well as the medium-band filters F210M and F360M where available. Coadded photometry from all existing images is used to construct our sample. The Wide epoch 2 images are essential for covering the dither gaps in short-wavelength filters from Wide epoch 1 images alone. 

Our spectroscopic analysis is based on the NIRSpec/MSA data from the Deep tier epochs 1--6, in which photometrically selected sources are targeted with PRISM observations at $R \sim 100$--300 over 0.6--5.3\,\micron. Each Deep epoch consists of four NIRSpec/MSA pointings, designed to accommodate approximately 800--900 MSA targets with different observing priorities assigned according to source brightness and photometry-based science categories \citep{ZhuangMingyang_2026_NEXUS_QR}. For example, all photometric LRD candidates from \citet{ZhuangMingyang_2026_NEXUS_LRD} are assigned Class 1, one of the highest-priority target classes. In addition, Class 4 comprises all targets with ${\rm F444W}<26$ mag. Although these targets have lower priority, they are expected by survey design to achieve nearly complete spectroscopic coverage after completion of the NEXUS program \citep{ShenYue_2024_NEXUS,ZhuangMingyang_2026_NEXUS_QR}. The observations adopt the \texttt{NRSIRS2RAPID} readout pattern with 43 groups, 1 integration, and a two-shutter slitlet nodding for each pointing, corresponding to an effective exposure time of 21.4 minutes per target. More technical details on survey design, observing schedule and data reduction are presented in \citet{ShenYue_2024_NEXUS} and \citet{ZhuangMingyang_2026_NEXUS_QR}. 

In total, the parent photometric catalog contains more than 180,000 objects over the full NEXUS-Wide footprint. The Deep-tier spectroscopic data used in this work include 5305 MSA spectra from epochs 1--6. These data will be released as part of NEXUS Data Release 1 (DR1; M.-Y. Zhuang et al., in preparation), and the DR1 LRD sample refers to the sample presented in this work.

\subsection{Photometric Selection of LRD Candidates}\label{sec:photo_LRD}
Since LRDs remain a phenomenological class without a universally accepted physical definition, we adopt in this work an observational definition that is commonly used in the literature for photometric LRDs. In general, LRDs are characterized by being compact and extremely red in the rest-optical, indicative of a dominant, unresolved red central component. In addition, many typical LRDs exhibit blue rest-UV continua, producing the characteristic “V-shaped” SED. Accordingly, the first two selection methods described in Sections~\ref{sec:V-color_selection} and~\ref{sec:V-slope_selection} are designed to identify such compact sources with V-shaped SEDs across rest-frame UV-optical. In some cases, however, the UV upturn is weak or poorly constrained. Therefore, in the third method described in Section~\ref{sec:red_selection}, we relax the rest-frame UV requirement and instead focus on selecting extremely red, compact objects.

We apply a combination of the three methods to select photometric LRD candidates over the currently available NEXUS-Wide imaging footprint, which covers approximately 382~arcmin$^2$. The subset of this sample falling within the central Deep tier is used for NIRSpec/MSA spectroscopic validation and will be discussed in Section~\ref{sec:spec_LRD}.

\subsubsection{Color selection for V-shaped SED}\label{sec:V-color_selection}

Photometric LRD candidates can be selected using the color selection method developed in previous studies to identify the V-shaped SED \citep{Greene_2024_Vshape_UNCOVER,Kokorev_2024_Vshape,Labbe_2025_Vshape_UNCOVER}. This selection is designed to isolate compact objects with blue rest-UV continua and red rest-optical colors, which together produce the characteristic V-shape spectral energy distribution.

Because NEXUS does not include F277W imaging, we estimate the F277W flux by fitting an LRD template, calibrated from the LRDs at $4<z<9$ of \citet{Kokorev_2024_Vshape}, to the F200W and F356W photometry. In addition, we take advantage of the additional medium-band coverage in NEXUS: F210M is used as a substitute for F200W if the latter is not available, and F360M as a substitute for F356W when F356W is unavailable.

\begin{figure*}
\epsscale{1.15}
\plotone{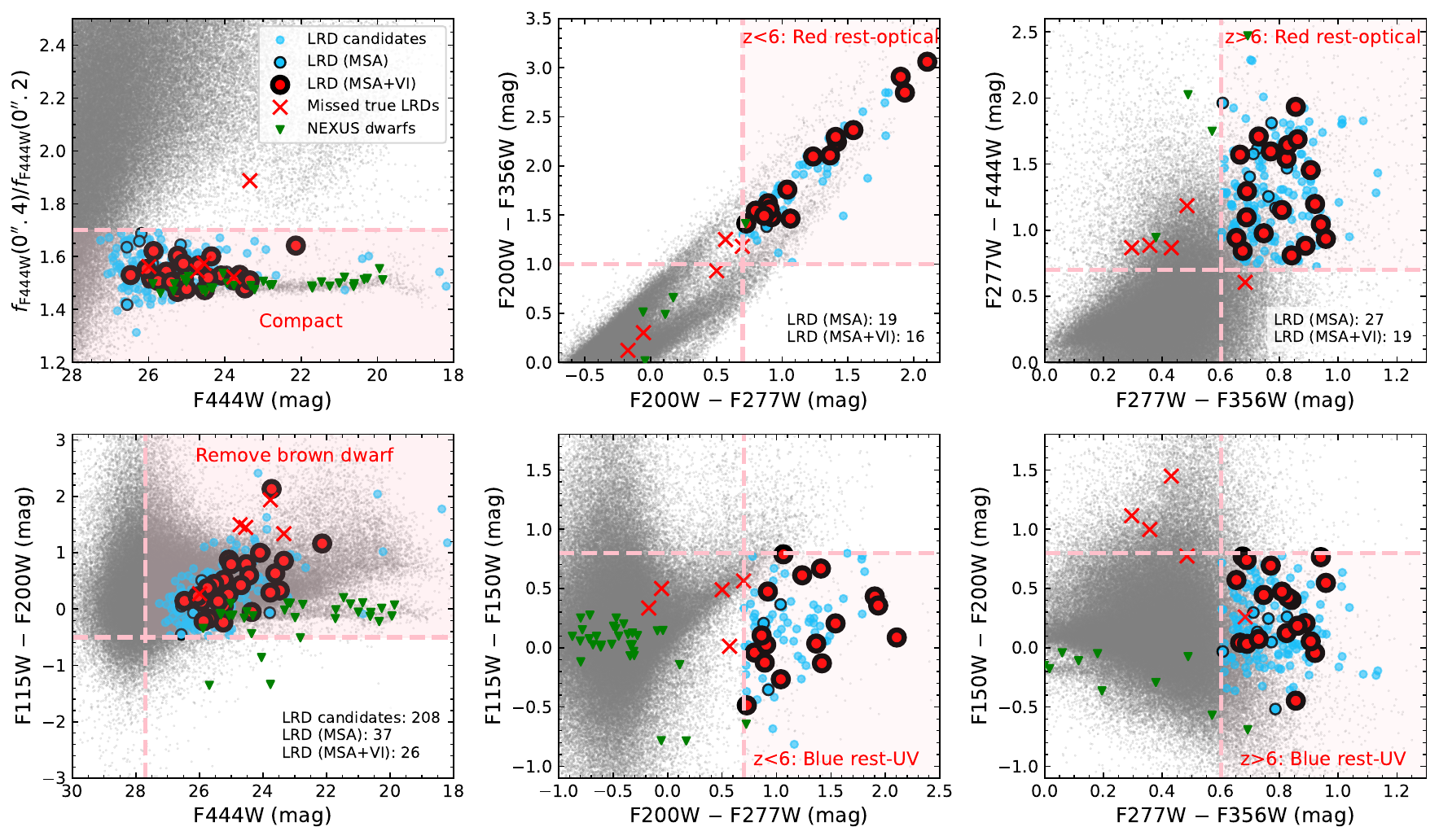}
\caption{
Color selection of LRD candidates exhibiting V-shaped SEDs, detailed in Section~\ref{sec:V-color_selection}. The upper panel in the leftmost column shows the compactness criterion, defined as $f_{\rm F444W}(0.4^{\prime\prime})/f_{\rm F444W}(0.2^{\prime\prime})$. The lower panel in the leftmost column illustrates the brown dwarf rejection cut, which is shown to be effective in Section~\ref{sec:spec_LRD} and Figure~\ref{fig:contamination}. The middle and rightmost columns present the color selections for the low-redshift ($z<6$) and high-redshift ($z>6$) samples, respectively, using red rest-frame optical and blue rest-frame UV colors to isolate sources with V-shaped SEDs. Gray points represent the full source population. Blue points denote photometrically selected LRD candidates, while blue points with black outlines indicate those with MSA spectroscopy. Red points with black outlines mark spectroscopically confirmed LRDs. Red crosses indicate spectroscopic LRDs missed by the photometric selection but visually recovered, primarily due to emission-line contamination in F200W and host-galaxy contamination affecting the compactness measurement. Green triangles denote the NEXUS brown dwarf sample (A. J. Burgasser et al., in preparation), shown as potential contaminants. The shaded pink regions indicate the adopted selection boundaries.
}\label{fig:Selection_Vshape}
\end{figure*}

We require candidates to be significantly detected in F444W, with ${\rm SNR}_{\rm F444W} > 14$ and ${\rm F444W} < 27.7$ mag, where ${\rm SNR}_{\rm F444W}$ is the signal-to-noise ratio in the F444W band. For the $z<6$ regime, we select sources with relatively blue rest-UV colors and red rest-optical colors using
\begin{equation}
\left\{
\begin{aligned}
\mathrm{F115W} - \mathrm{F150W} &< 0.8, \\
\mathrm{F200W} - \mathrm{F277W} &> 0.7, \\
\mathrm{F200W} - \mathrm{F356W} &> 1.0.
\end{aligned}
\right.
\end{equation}

For the $z>6$ regime, the selection shifts to longer wavelengths, and candidates are required to satisfy
\begin{equation}
\left\{
\begin{aligned}
\mathrm{F277W} - \mathrm{F356W} &> 0.6, \\
\mathrm{F277W} - \mathrm{F444W} &> 0.7, \\
\mathrm{F150W} - \mathrm{F200W} &< 0.8.
\end{aligned}
\right.
\end{equation}
When applying the color criteria, we require each source to be detected at $>3\sigma$ in at least one band of each color term. For non-detections, we adopt the $2\sigma$ upper limit, but only when the brighter band in the corresponding color is detected at $>3\sigma$. In Figure~\ref{fig:Selection_Vshape}, the color selection criteria for the $z<6$ regime are illustrated in the middle column, whereas those for the $z>6$ regime are shown in the right column. These criteria select sources with red rest-optical colors and blue rest-UV continua.

In addition to these color criteria, sources are required to be spatially unresolved in F444W. Following \citet{Kokorev_2024_Vshape}, we quantify compactness using the aperture flux ratio
$f_{\rm F444W}(0\farcs4)/f_{\rm F444W}(0\farcs2) < 1.7$,
where $f_{\rm F444W}(0\farcs4)$ and $f_{\rm F444W}(0\farcs2)$ are the fluxes measured within $0\farcs4$ and $0\farcs2$ apertures, respectively. Finally, LRD candidates are required to satisfy the brown dwarf rejection criterion $\mathrm{F115W} - \mathrm{F200W} > -0.5$, as shown in the bottom left panel of Figure~\ref{fig:Selection_Vshape}.

\subsubsection{Slope selection for V-shaped SED}\label{sec:V-slope_selection}

\begin{figure*}
\epsscale{1.03}
\plotone{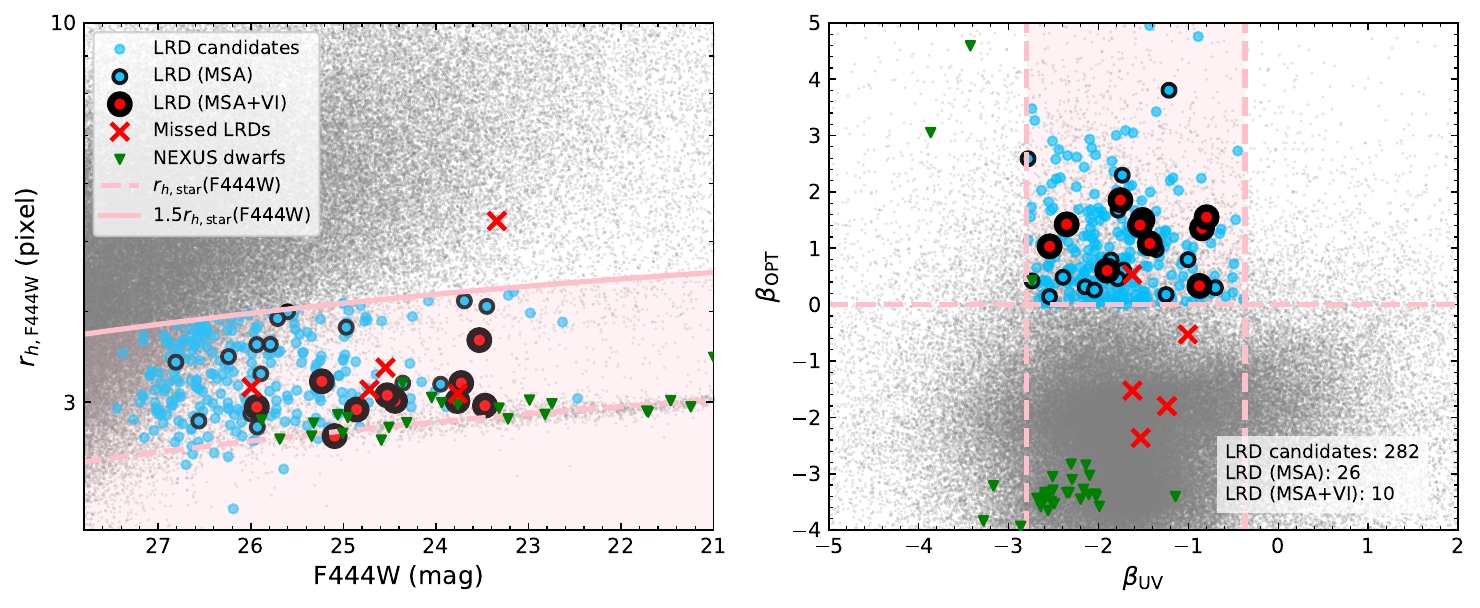}
\caption{Slope selection of LRD candidates exhibiting V-shaped SEDs, detailed in Section~\ref{sec:V-slope_selection}. Left: half-light radius in F444W, $r_{h, \rm F444W}$ versus F444W magnitude for the bright sample (${\rm SNR}_{\rm F444W} > 12$). The dashed and solid pink curves show the stellar half-light radius and 1.5 times that value, respectively; the shaded region marks the compact-source selection. Right: $\beta_{\rm OPT}$ versus $\beta_{\rm UV}$ for the same sample. The shaded region indicates the adopted V-shaped SED selection, defined by $-2.8 < \beta_{\rm UV} < -0.37$ and $\beta_{\rm OPT} > 0$. The symbol coding is the same as in Figure~\ref{fig:Selection_Vshape}.}
\label{fig:Selection_Slope}
\end{figure*}

\citet{Kocevski_2025_Slope} identified LRD candidates with V-shaped SEDs using a photometric slope-based selection method, which was adopted in our earlier NEXUS LRD paper based on NIRCam/WFSS spectroscopy \citep{ZhuangMingyang_2026_NEXUS_LRD}. This method has the following photometric selection criteria:
\begin{equation}
\left\{
\begin{aligned}
{\rm SNR}_{\rm F444W} &> 12, \\
-2.8 < \beta_{\rm UV} &< -0.37, \\
\beta_{\rm opt} &> 0, \\
r_{h, \rm F444W} &< 1.5 \times r_{h,\mathrm{star}},
\end{aligned}
\right.
\end{equation}
where we define the continuum slope $\beta$ through $f_\lambda \propto \lambda^\beta$. $\beta_{\rm UV}$ and $\beta_{\rm opt}$ are measured from multi-band photometry in the rest-frame UV and optical, respectively. The requirement on $\beta_{\rm UV}$ selects sources with blue rest-frame UV continua, while the condition on $\beta_{\rm opt}$ selects sources with red rest-frame optical continua. The quantity $r_{h,\rm F444W}$ denotes the half-light radius in the F444W band, and $r_{h, \rm star}$ represents the stellar half-light radius used to define the compactness criterion. Specifically, we fit a second-order polynomial to the stellar locus to derive a magnitude-dependent size threshold. We then select compact sources with sizes smaller than 1.5 times the stellar size at a given magnitude, as illustrated in Figure~\ref{fig:Selection_Slope}.

\begin{deluxetable*}{Ccc}
\caption{Filter sets for continuum slope estimation}
\label{tab:filters}
\tablehead{
\colhead{Photometric Redshift} & \colhead{Rest-UV Slope ($\beta_{\rm UV}$)} & \colhead{Rest-optical Slope ($\beta_{\rm opt}$)}
}
\startdata
$2 \leq z < 3.25$     & F090W, F115W                    & F150W, F200W (F210M) \\
$3.25 \leq z < 4.75$  & F090W, F115W, F150W            & F200W (F210M), F356W (F360M) \\
$4.75 \leq z < 8$     & F115W, F150W, F200W (F210M)    & F356W (F360M), F444W \\
$z \geq 8$            & F150W, F200W (F210M)           & F356W (F360M), F444W \\
\enddata
\tablecomments{Bands in parentheses are used as substitutes when the corresponding primary bands are not detected or not covered by the data: F210M replaces F200W, and F360M replaces F356W.}
\end{deluxetable*}

\begin{figure*}
\epsscale{1.03}
\plotone{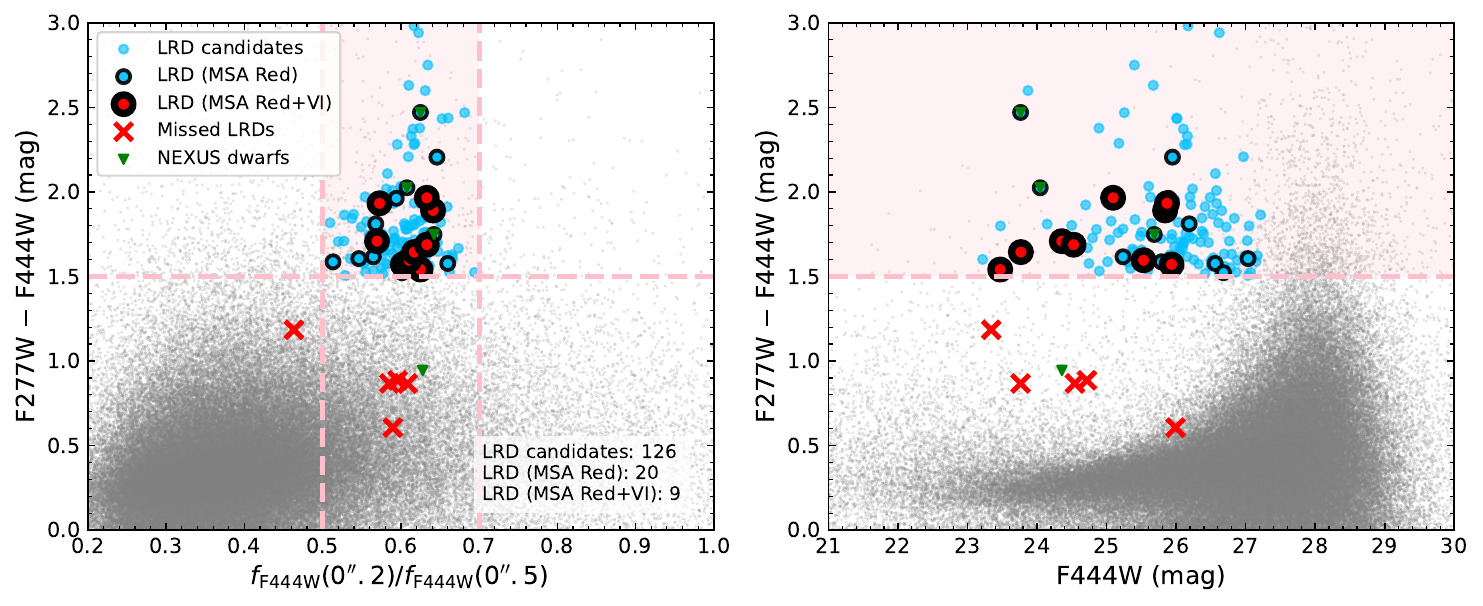}
\caption{Color selection of LRD candidates exhibiting very red SEDs, detailed in Section~\ref{sec:red_selection}. Left: F277W$-$F444W color versus $f_{\rm F444W}(0\farcs2)/f_{\rm F444W}(0\farcs5)$. The shaded region and dashed lines indicate the adopted selection criteria, requiring $0.5 < f_{\rm F444W}(0\farcs2)/f_{\rm F444W}(0\farcs5) < 0.7$ and F277W$-$F444W $> 1.5$. Right: F277W$-$F444W color versus F444W magnitude. The symbol coding is the same as in Figure~\ref{fig:Selection_Vshape}. }
\label{fig:Selection_Red}
\end{figure*}

To estimate $\beta_{\rm UV}$ and $\beta_{\rm opt}$, we use the photometric redshifts, ($z_{\rm phot}$) derived using EAZY \citep{Brammer_2008_EAZY} with the \texttt{agn\_blue\_sfhz\_13} template set. In addition to the NIRCam photometry from NEXUS, we include Subaru Hyper Suprime-Cam (HSC) photometry from the HEROES survey \citep{Taylor_2023_HEROES}, consisting of five broad bands ($grizy$) and two narrow bands (NB816 and NB921), in the photo-$z$ estimation. Depending on the source photometric redshift, we choose the appropriate filter sets for continuum-slope measurements, accounting for bandpass shifting with redshift. In principle, this method should allow the identification of LRDs across a broader redshift range ($z \sim 2-11$) than simple color–color selection. The adopted filter sets in four photometric-redshift bins are listed in Table~\ref{tab:filters}.

\subsubsection{Color selection for very red SEDs}\label{sec:red_selection}

\citet{Akins_2025_Cosmos_Red} selected compact sources in F444W, but did not impose a blue rest-frame UV criterion. Instead, they focused on the reddest objects by requiring
\begin{equation}
\left\{
\begin{aligned}
{\rm SNR}_{\rm F444W} &> 12, \\
{\rm SNR}_{\rm F277W} &> 3, \\
{\rm F277W} - {\rm F444W} &> 1.5, \\
0.5 < f_{\rm F444W}(0\farcs2)/f_{\rm F444W}(0\farcs5) &< 0.7,
\end{aligned}
\right.
\end{equation}
where $ f_{\rm F444W}(0\farcs2)/f_{\rm F444W}(0\farcs5)$ denotes the compactness measured from the ratio of fluxes within the $0\farcs2$ and $0\farcs5$ apertures. Note that we have followed the same definitions of source compactness for each selection method in previous studies. Although these compactness definitions are not identical, the impact is minimal on the final selection of LRDs. 

Figure~\ref{fig:Selection_Red} shows the implementation of this selection using NEXUS photometry: the left panel presents the color--compactness plane used to identify compact red sources, while the right panel shows the corresponding ${\rm F277W}-{\rm F444W}$ color as a function of F444W magnitude. Compared with other photometric selection methods, the ${\rm F277W} - {\rm F444W} > 1.5$ criterion targets extremely red sources and therefore preferentially selects objects at $z \gtrsim 5$. Although emission lines from extreme emission-line galaxies can artificially boost the F444W flux and mimic red colors, the stringent red-color cut adopted here helps to largely mitigate this contamination. On the other hand, because this method focuses only on the reddest part of the population and does not require a UV upturn, it may still include some dusty compact high-redshift objects that are not typical LRDs.

\subsection{Spectroscopic LRDs from MSA}\label{sec:spec_LRD}

\begin{figure}
\epsscale{1.2}
\plotone{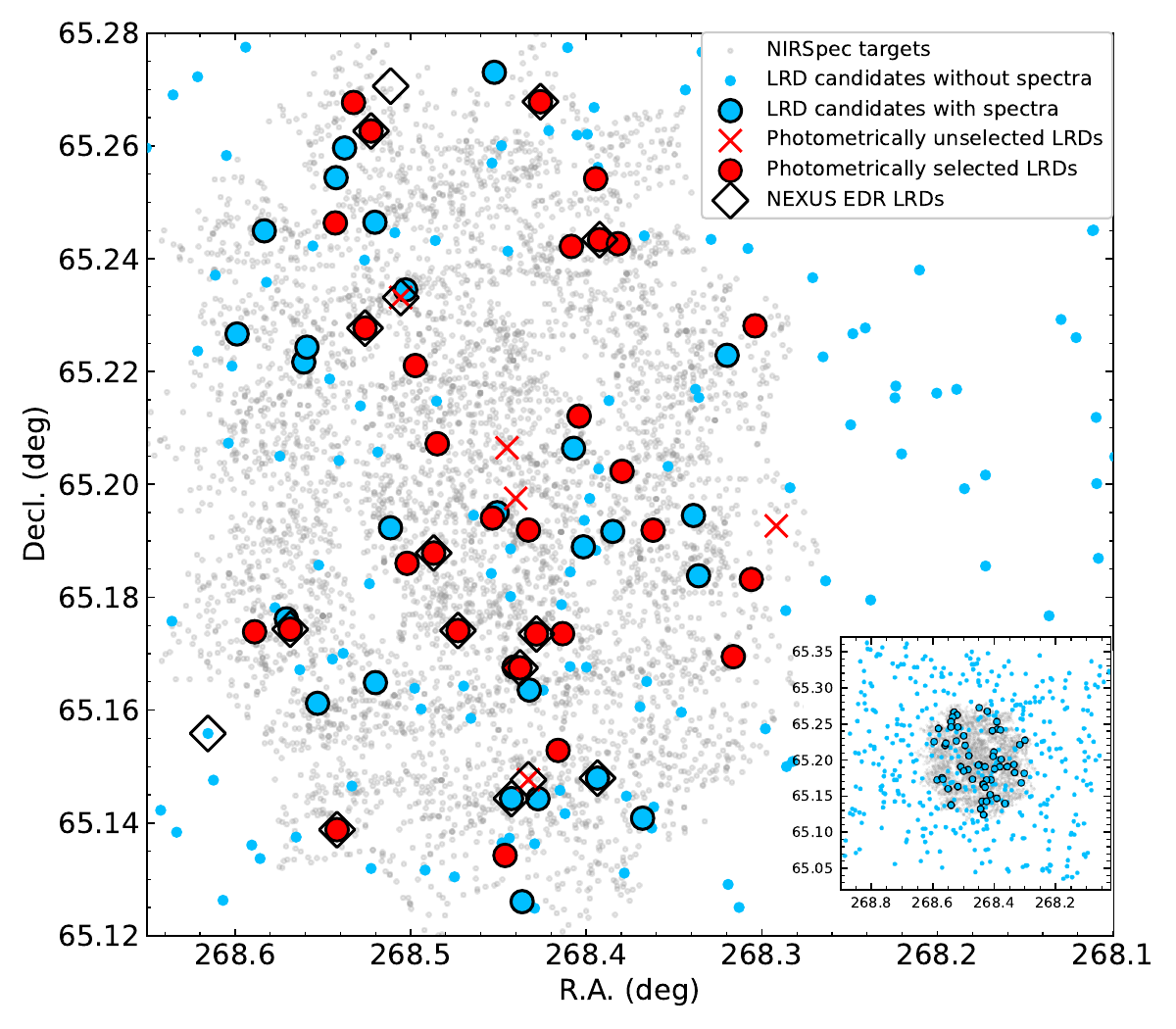}
\caption{
Spatial distribution of the LRD population in the NEXUS Deep field. Small gray points show the positions of all NIRSpec MSA targets from the multi-epoch deep spectroscopic observations. Photometric LRD candidates are shown in blue, with objects lacking valid spectra plotted as small blue points and those with valid spectroscopic coverage shown as larger blue circles with black outlines. Spectroscopically identified LRDs in our sample are overplotted in red: photometrically selected LRDs are shown as filled red circles with black outlines, while photometrically unselected LRDs are marked by red crosses. Previously identified NEXUS EDR LRDs are shown as open black diamonds. The main panel zooms in on the central spectroscopically covered region, and the inset shows the broader field.
} 
\label{fig:LRD_spatial}
\end{figure}

We identify 463 photometric LRD candidates using the three selection methods described above. We then perform spectroscopic inspection for a subset of this sample with NIRSpec MSA observations in the NEXUS Deep field. As shown in Figure~\ref{fig:LRD_spatial}, the main panel presents a zoomed-in view of the Deep field covered by the MSA observations, with the observed sources marked by gray dots, while the inset shows the full NEXUS-Wide field. Among the 463 photometric candidates, 57 have MSA spectra from the first 6 NEXUS-Deep epochs. 

To more robustly characterize the V-shaped continua of LRDs, we measure spectral slopes directly from the PRISM spectra. We fix the continuum break wavelength at the Balmer limit and fit the spectra on either side with independent power laws of the form $f_\lambda = a\lambda_{\rm rest}^{\beta}$ using nonlinear least-squares optimization. By default, the rest-UV fitting window spans 1200~\AA\ to the Balmer limit, while the rest-optical window extends from the Balmer limit to 7000~\AA. Strong emission lines are masked during the fitting. For a few sources whose continuum breaks are not located near the Balmer limit, we manually adjust the fitting setup.

\begin{figure}
\epsscale{1.15}
\plotone{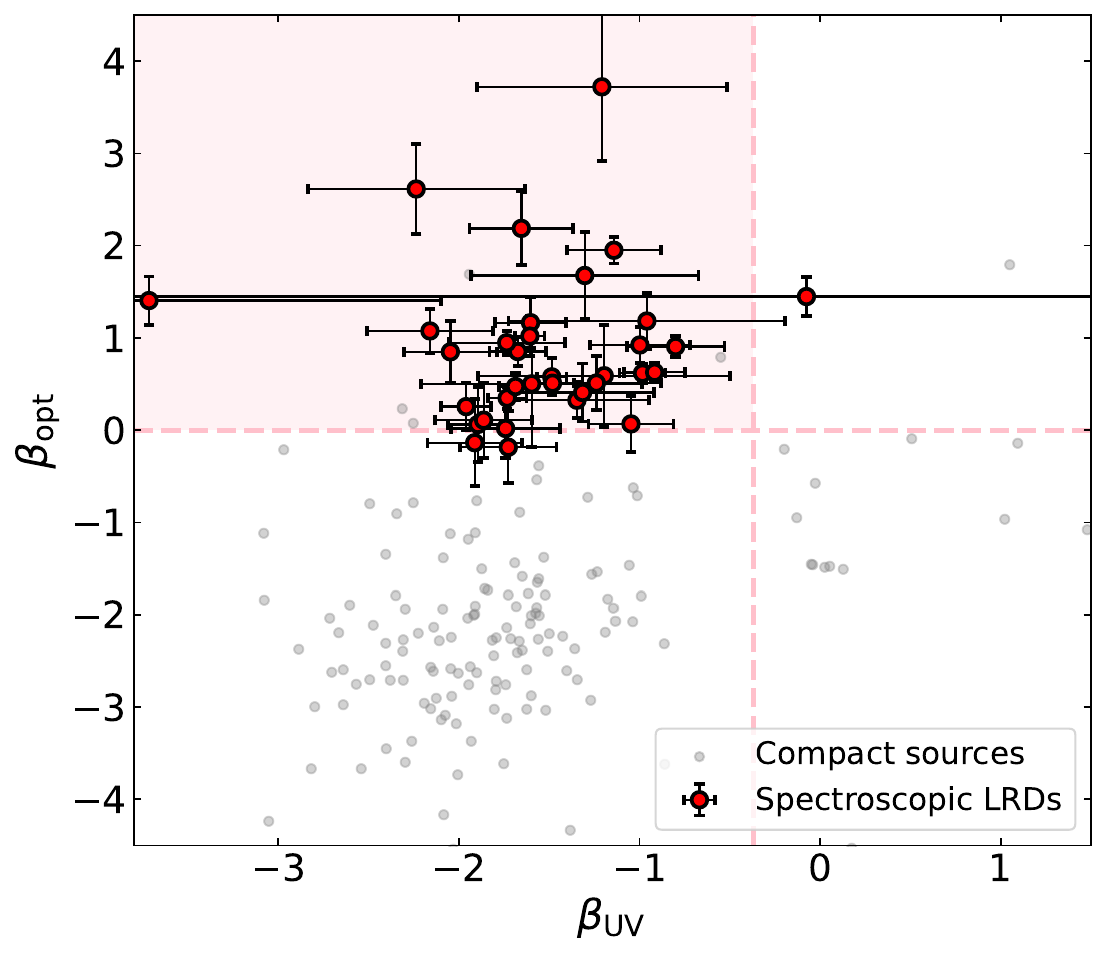}
\caption{
Spectral-slope selection of spectroscopic LRDs exhibiting V-shaped continua. Grey points represent the $z>2$ compact sources, while red circles mark the spectroscopic LRDs. The shaded pink region indicates the adopted V-shaped continuum selection, defined by $\beta_{\rm UV}< -0.37$ and $\beta_{\rm opt}>0$.
} 
\label{fig:spectral_slopes}
\end{figure}

We adopt a commonly used definition of spectroscopic LRDs, requiring a V-shaped continuum together with compact morphology \citep[e.g.,][]{Hviding_2025}. We first select compact objects at $z>2$ that satisfy at least one of the three compactness criteria used in Section~\ref{sec:photo_LRD}. We then further require $\beta_{\rm UV}< -0.37$ and $\beta_{\rm opt}>0$. After careful visual inspection, we exclude four objects (satisfying these spectral slope criteria) with ${\rm F444W}>26$~mag whose low continuum SNR makes their nature difficult to constrain. In addition, we manually recover three objects that lie close to the color-selection boundary. As shown in Figure~\ref{fig:spectral_slopes}, this procedure yields a spectroscopic LRD sample of 36 objects.

This criterion follows a similar logic to the photometric selection, while spectroscopy allows us to distinguish intrinsic continuum shapes from those impacted by strong emission lines falling into broad photometric bandpasses. Based on a qualitative visual classification, we define the canonical LRDs as “V-shaped LRDs.” In cases where the rest-optical continuum turns over toward longer wavelengths past the peak, forming a distinct “laid-down S” shape, we classify them as ``S-shape LRDs''. These two subclasses constitute the most secure LRD population, exhibiting the clearest spectral signatures of the LRD phenomenon.

We also consider a more tentative class of LRD candidates, referred to as ``L-shape LRDs'', which show a weaker or absent Balmer break but still exhibit a red rest-optical continuum compared to typical blue AGNs. These objects likely lie near the boundary of the photometric selection criteria, and similar cases have been reported in the literature as LRDs \citep[e.g., MoM 149501 and RUBIES-UDS 167741;][]{Hviding_2025,deGraaff_2025}. To minimize contamination from reddened AGN, we further require that L-shape LRDs lack the strong UV emission lines commonly seen in classical AGN. Our spectroscopic sample is therefore dominated by S-shape and V-shape LRDs, while retaining L-shape systems to capture the diversity of spectral properties and enable a more comprehensive exploration of the LRD population.

\begin{figure}
\epsscale{1.0}
\plotone{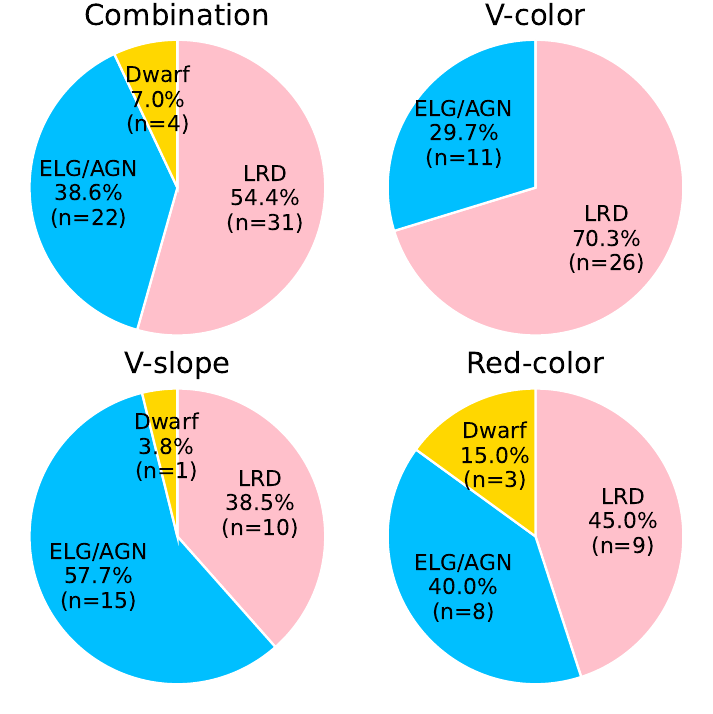}
\caption{
Breakdown of source types of the photometrically selected LRD samples after spectroscopic inspection. Each pie chart shows the fractions of spectroscopically confirmed LRDs, ELGs/AGNs, and dwarf stars among the photometric LRD candidates. 
} 
\label{fig:contamination}
\end{figure}

Figure~\ref{fig:contamination} shows the breakdown of the photometrically selected LRD candidates after spectroscopic inspection. Each pie chart illustrates the fractions of spectroscopically confirmed LRDs, emission-line galaxies (ELG)/AGN, and dwarf stars among the candidates that have MSA spectra. Overall, the dominant source of contamination comes from strong emission-line objects. We do not distinguish between ELGs and AGNs here, as such a separation is often ambiguous at high redshift and at the faint end in PRISM spectra. Strong emission lines in these objects significantly boost the broadband fluxes, thereby producing red rest-optical colors, while their intrinsic rest-UV continuum may remain blue. Although these photometric selection methods are already designed to mitigate this effect, they cannot completely eliminate such contamination. The ELG/AGN contamination fraction is highest in the V-slope selected sample, as shown in the lower left pie chart of Figure~\ref{fig:contamination}. This is likely because the V-slope method relies on robust photometric redshifts. For sources with actual redshifts near the boundaries of the redshift bin, the filter set adopted for slope estimation may be inadequate, leading to misidentified LRD candidates. We emphasize that this is not an inherent problem with the selection technique itself, but programs with less reliable photometric redshifts might see a higher contamination rate and/or lower completeness of LRD selection using the V-slope method.

Another important contaminant is brown dwarfs. The upper right pie chart shows that the V-color selected sample has no dwarf contamination, consistent with the brown dwarf rejection criterion included in that selection. Specifically, we apply a simple color cut, $\mathrm{F115W} - \mathrm{F200W} > -0.5$, and its performance suggests that this criterion is effective, at least for the NEXUS sample. In contrast, the red-color method shows the highest fraction of brown dwarf contamination, likely because we do not impose any constraint on the rest-UV spectral shape. \citet{Akins_2025_Cosmos_Red} applied a grid-fitting procedure using dwarf templates to remove this contamination and reported a dwarf contamination rate of roughly 24\%, broadly consistent with our result.

\begin{figure*}
\epsscale{1.1}
\plotone{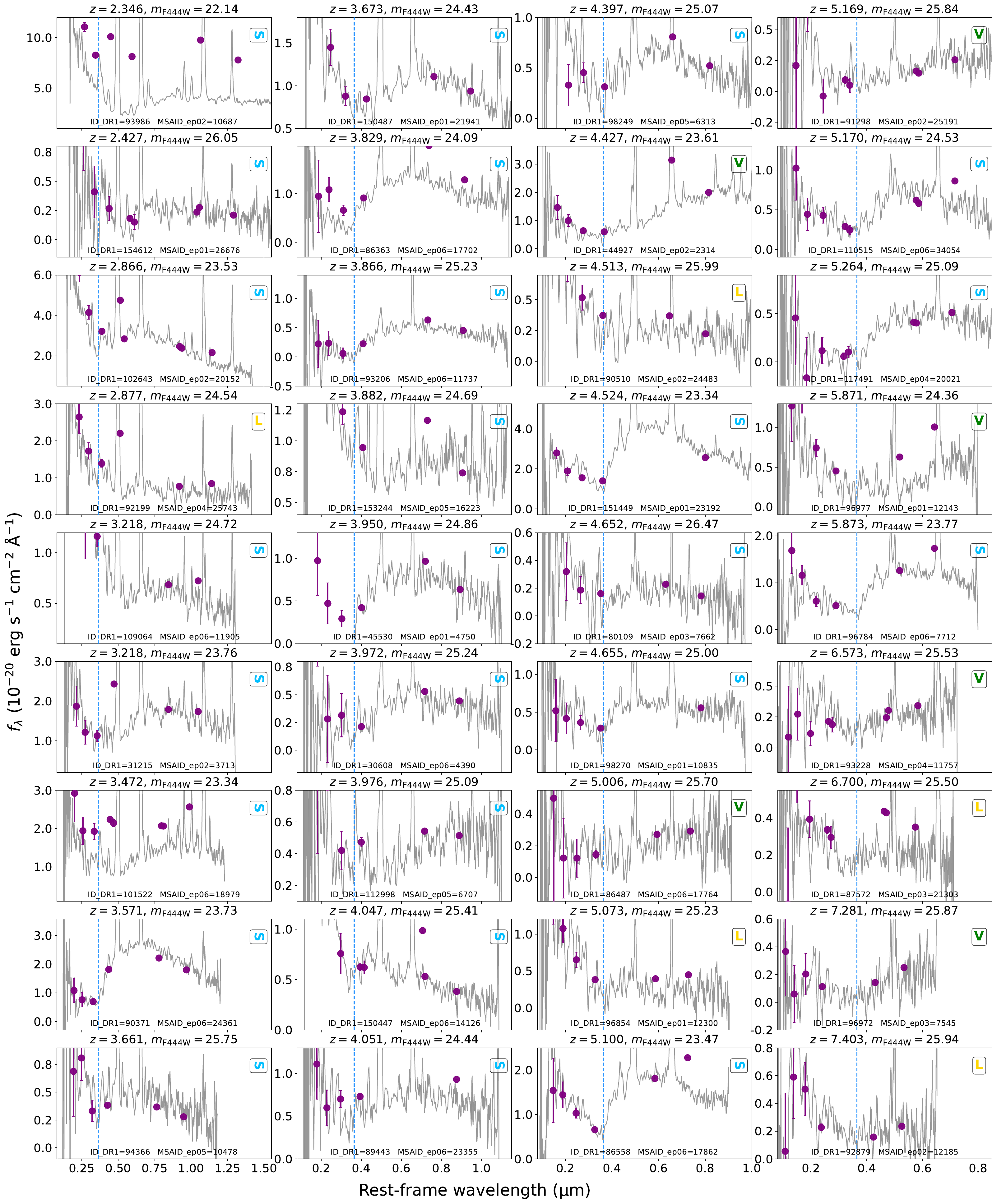}
\caption{
Rest-frame spectral gallery of the spectroscopically confirmed LRD sample with the order of increasing redshift. For each source, the gray curve shows the smoothed JWST/NIRSpec prism spectrum, while the purple points indicate the NIRCam photometry. The blue dashed line marks the Balmer break at $3646$\,\AA. The upper-right corner of each panel indicates the SED-shape subclass, where ``S'', ``V'', and ``L'' denote the S-shape, V-shape, and L-shape LRDs, respectively.
} 
\label{fig:Spec_gallery}
\end{figure*}

The spectroscopic sample of 36 LRDs includes five objects that are missed by our photometric selection, as shown in Figure~\ref{fig:Spec_gallery}. Three of them, ID-92199 ($z=2.887$, photo-$z=3.37$), ID-109064 ($z=3.218$, photo-$z=3.39$), and ID-31215 ($z=3.223$, photo-$z=3.40$), are missed because \Hb{}+\OIII{} contamination boosts the F200W flux, preventing the V-shaped continuum from being correctly identified by the V-color method. For the V-slope method, all three sources have photometric redshifts greater than 3.25, for which F200W and F356W are used to measure the optical slope (see Table~\ref{tab:filters}). The line-boosted F200W flux therefore causes these sources to fail the V-slope selection. ID-101522 ($z=3.471$) is missed because of both emission-line contamination and host-galaxy emission affecting the compactness measurement. Its compact core is identified as an LRD through more careful image decomposition \citep{ZhuangMingyang_2026_NEXUS_LRD}. ID-90510 ($z=4.513$, photo-$z=5.75$) lies slightly outside the V-color selection boundaries, and its incorrect photometric redshift causes it to fail the V-slope selection.

\begin{figure*}
\epsscale{1.15}
\plotone{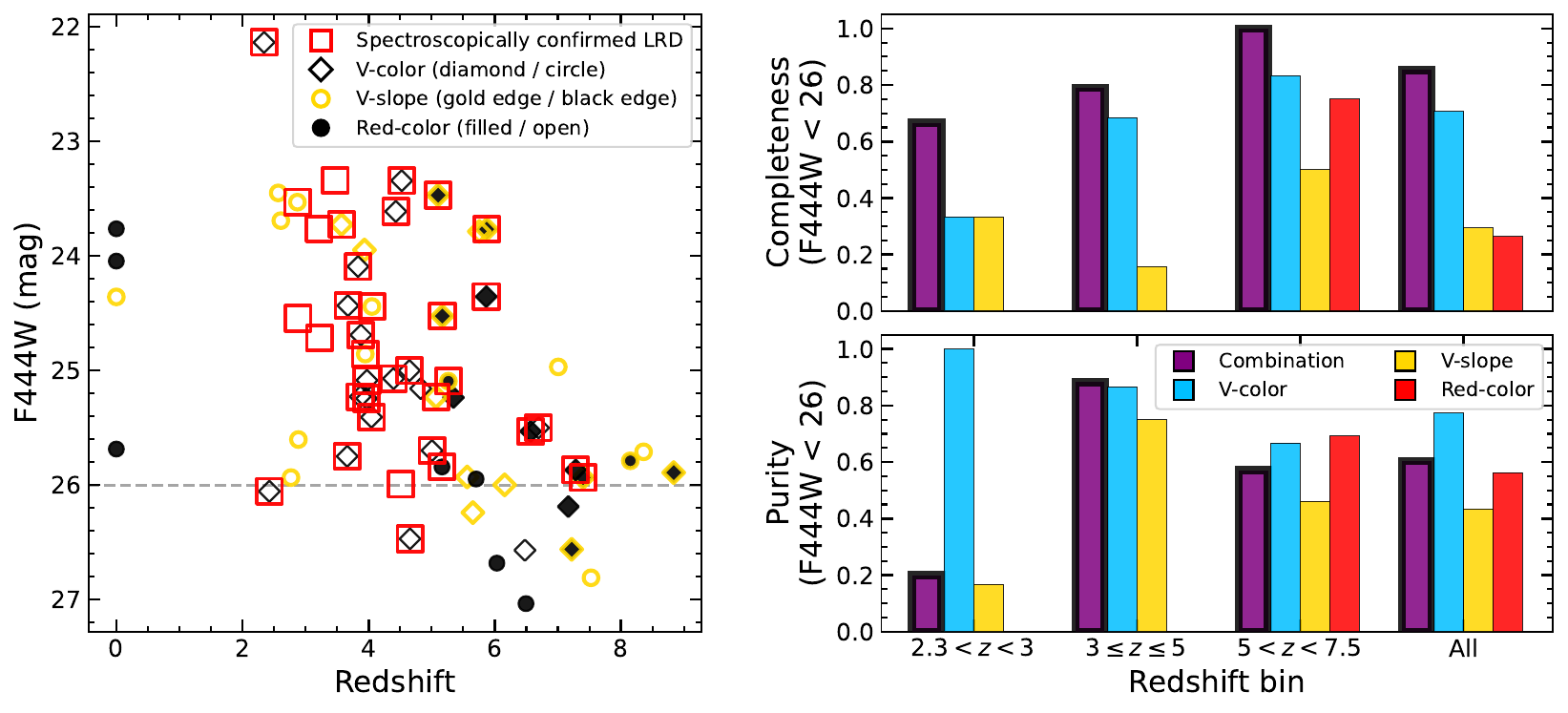}
\caption{
Left: spectroscopic and photometric LRD samples in the F444W magnitude--redshift plane. The sample includes 36 spectroscopically confirmed LRDs, marked by red square outlines, and 57 photometrically selected LRD candidates with MSA spectroscopic observations. The three photometric selection methods are encoded simultaneously in the symbol style: diamond and circular markers indicate the V-color selection being satisfied or not, respectively; gold and black marker outlines indicate the V-slope selection being satisfied or not, respectively; and filled and open markers indicate the Red-color selection being satisfied or not, respectively. Right: completeness (top) and purity (bottom) of the photometric LRD selection as a function of redshift, computed using only sources with ${\rm F444W} < 26$ mag. In each redshift bin, four bars show the values for the V-color method (blue), the V-slope method (yellow), the Red-color method (red), and the combination of the three methods (purple).
}\label{fig:Selection_mz}
\end{figure*}

\subsection{Completeness and purity}\label{sec:comp}

In total, our final catalog contains 36 spectroscopically confirmed LRDs based on the MSA data, 31 of which are recovered by our photometric selection. Their rest-frame spectral gallery in order of increasing redshift is shown in Figure \ref{fig:Spec_gallery}. The corresponding photometric selection method(s) and SED-shape subclass for each source are listed in Table~\ref{tab:catalog}. The overall completeness of our combined photometric selection, based on all three methods, is approximately $31/36 \sim 86\%$. We note that two NEXUS EDR LRDs from \citet{ZhuangMingyang_2026_NEXUS_LRD}, indicated by black diamonds in Figure~\ref{fig:LRD_spatial}, do not yet have MSA observations and are thus excluded from the present estimates of completeness and purity.

The 31 spectroscopic LRDs among the 57 photometric LRDs with MSA spectra, highlighted as filled red circles in Figure~\ref{fig:LRD_spatial}, indicate that the overall purity of our photometric selection is $31/57 \sim 54\%$. Restricting the sample to sources with ${\rm F444W}<26$ mag, for which the spectroscopic coverage is more complete  \citep{ZhuangMingyang_2026_NEXUS_QR}, yields a slightly higher purity of $29/48 \sim 60\%$.

Figure~\ref{fig:Selection_mz} shows the distribution of the LRD sample in the F444W magnitude--redshift plane (left) and the performance of each photometric selection method. In the left panel, the 36 spectroscopically confirmed LRDs, marked by red square outlines, span a broad range in both redshift and F444W magnitude. Also shown are the 56 photometrically selected LRD candidates (four spectroscopically confirmed brown dwarfs are at $z=0$) with MSA spectroscopic observations, with different symbols and colors for the V-color, V-slope, and Red-color selection, respectively.

The right panels of Figure~\ref{fig:Selection_mz} show the completeness and purity of each photometric LRD selection in different redshift bins. These quantities are computed only using sources with ${\rm F444W} < 26$~mag, for which the NEXUS Deep-tier observations provide more complete spectroscopic coverage by design \citep{ShenYue_2024_NEXUS,ZhuangMingyang_2026_NEXUS_QR}. The V-color, V-slope, and red-color selections exhibit different redshift dependences. At $z\sim3$, as discussed in Section~\ref{sec:spec_LRD}, emission-line contamination in F200W significantly degrades the performance of all three photometric selection methods. In addition, the number density of LRDs at $z<3$ likely declines rapidly relative to that at higher redshifts, so the small sample size makes the estimation in this bin less statistically robust.

In the $3<z<5$ bin, both the V-color and V-slope methods achieve high purity, exceeding 80\%, although only the V-color method also maintains high completeness ($>60\%$). The lack of F277W coverage and the strict photometric-redshift requirement likely contribute to the low completeness of the V-slope selection. At $z>5$, the Red-color method performs efficiently, at least for the NEXUS data. Compared with the V-color criterion of ${\rm F277W} - {\rm F444W} > 0.7$, the red-color method imposes a redder threshold of ${\rm F277W} - {\rm F444W} > 1.5$, which largely mitigates contamination from emission-line galaxies and yields a higher purity even without requiring a UV upturn. For the subsample at $z>6.5$, however, we have only three spectroscopically confirmed LRDs. The reduced performance in this regime likely arises from both the depth limit of the NEXUS data and the interpolation of F277W photometry. This interpolation is expected to be most biased in this redshift range because the Balmer break moves into F277W, where the true flux should lie near the SED minimum rather than near the midpoint between F200W and F444W fluxes.

As mentioned in Section~\ref{sec:NEXUS}, MSA targets in different classes have different observing priorities and therefore different spectroscopic completeness in the current six epochs. We therefore perform a more careful purity analysis that accounts for the MSA target priorities. Specifically, we estimate the purity separately for different target classes and derive a weighted purity using the fraction of photometric candidates in each class within the MSA footprint. This yields a purity estimate similar to our fiducial result, mainly because the overall spectroscopic completeness of LRD targets is already high. For photometric LRD targets with ${\rm F444W}<26$ mag, the overall spectroscopic completeness already reached 73\%.

We compare our estimates with those of \citet{Hviding_2025}, who used a spectroscopic LRD sample from RUBIES to assess the accuracy and completeness of similar photometric selection methods to those adopted in Section~\ref{sec:photo_LRD}. Their analysis was based on cross-matching the photometric LRD catalogs of \citet{Kokorev2024_LRD} and \citet{Kocevski2024_LRD} with the RUBIES spectroscopic sample. In contrast, our photometric selection and spectroscopic follow-up are both based on the same NEXUS parent catalog. This difference in sample construction may partly contribute to the slightly higher completeness found in this work, but our result is broadly consistent with that of \citet{Hviding_2025}. For the purity estimate, the comparison is more sensitive to the adopted definitions of confirmation and contaminants. In this work, we count all spectroscopically uncertain photometric candidates as contaminants (which remained in the denominator), whereas \citet{Hviding_2025} removed uncertain cases from their purity calculation (both in numerator and denominator). This difference naturally leads to a lower inferred purity in our analysis compared to the value of $\sim 90\%$ reported by \citet{Hviding_2025}. If uncertain cases are retained in the denominator, the corresponding purity fraction in \citet{Hviding_2025} is also around 60\%, consistent with our estimate.

In summary, over the full redshift range, the combined selection reaches a completeness of nearly 85\% and a purity of about 60\%, indicating that combining multiple photometric methods can yield a relatively complete and robust LRD sample. In particular, in the $3 \leq z < 5$ bin, the combined selection performs especially well, achieving both completeness and purity above 80\%. Although the Red-color method is not designed to select sources at $z<5$, it plays an important role at $z>5$ in the combined photometric selection.

\section{Results} \label{sec:results}

\subsection{Spectral Properties of NEXUS MSA LRDs} \label{sec:specfit}
\begin{figure*}
\epsscale{1.15}
\plotone{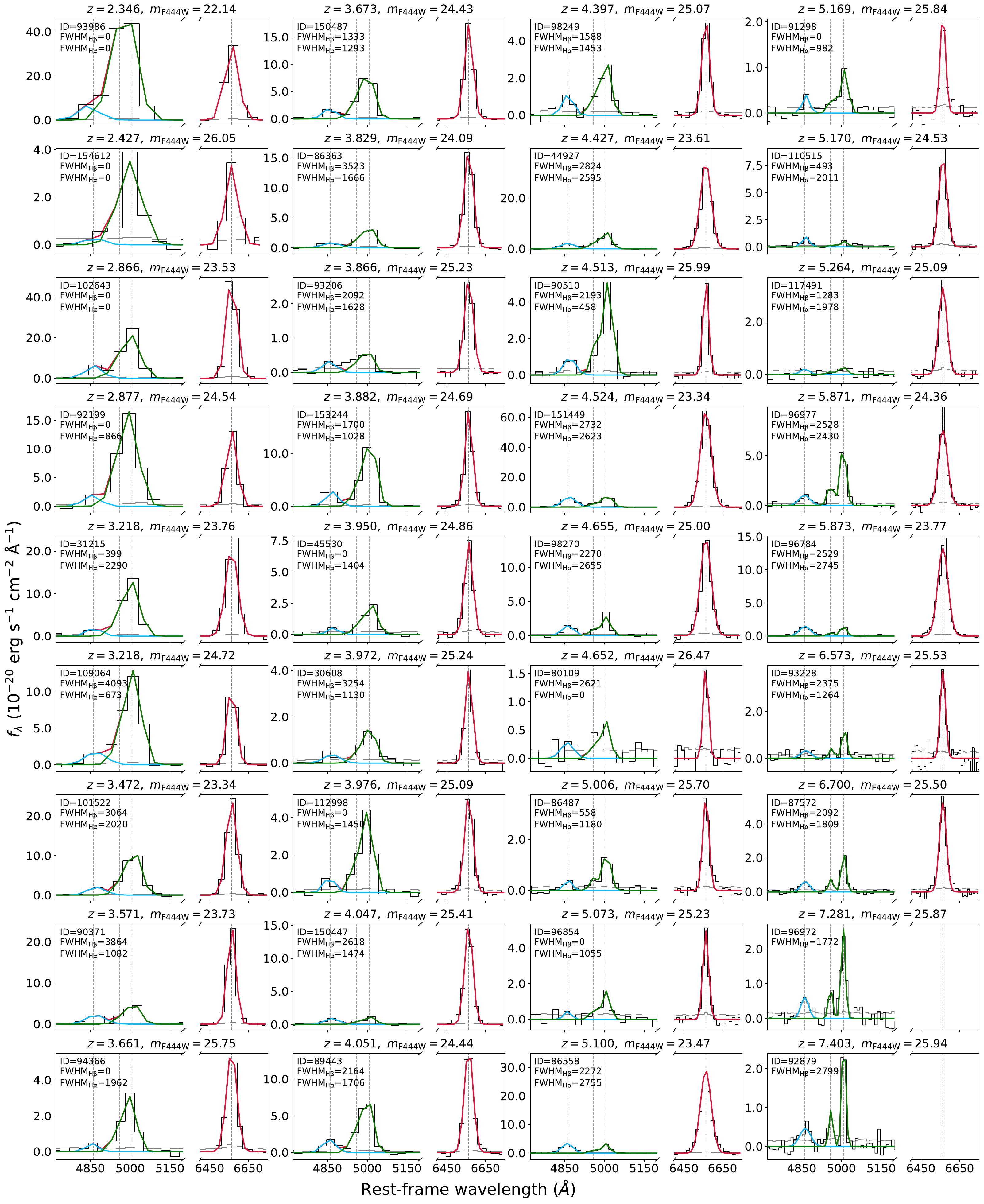}
\caption{
\Hb+\OIIIab{} and \Ha{} spectral fits for the spectroscopic LRD sample. In each panel, the black curve shows the continuum-subtracted rest-frame spectrum, the gray curve shows the 1$\sigma$ uncertainty, and the red curve shows the total best-fit model. The left and right sides of each panel display the \Hb+\OIII{} and \Ha{} regions, respectively. In the \Hb+\OIII{} region, the fitted \Hb{} and \OIII{} components are shown in blue and green, respectively. The measured FWHM values of \Hb{} and \Ha{} are labeled in each panel and have been corrected for instrumental broadening using the estimate from \citet{deGraaff_2024}. A value of zero indicates that the line is unresolved at the prism resolution.
}\label{fig:fitting}
\end{figure*}

In the following analysis, we focus on the 36 spectroscopically confirmed LRDs with MSA spectra. We first characterize their spectral properties using simple spectroscopic fitting. For each source, we fit the \Ha\ and \Hb+\OIII\ regions separately in the rest frame as shown in Figure~\ref{fig:fitting}. Within each fitting window, the local continuum is modeled as a power law and fitted simultaneously with the emission-line components.

For the \Ha{} region, we fit the spectrum over a rest-frame window of 6350--6750\,\AA. We adopt a single-Gaussian model, in which \Ha{} is represented by one Gaussian component with a free centroid, amplitude, and FWHM. The lower bound on the line width is set by the instrumental resolution, while the upper bound is taken to be the larger of \(2500\ {\rm km\ s^{-1}}\) and \(1.5\times{\rm FWHM}_{\rm inst}\).


\begin{rotatetable*}
\begin{deluxetable*}{lcccccccccccc}
\label{tab:catalog}
\tabletypesize{\scriptsize}
\tablecaption{Spectroscopic LRD Sample}
\tablehead{
\colhead{QDR ID} &
\colhead{R.A.} &
\colhead{Decl.} &
\colhead{$z_{\rm spec}$} &
\colhead{$m_{\rm F444W}$} &
\colhead{FWHM} &
\colhead{$\log\,L_{\rm H\alpha,broad}$} &
\colhead{$M_{\rm UV}$} &
\colhead{Shape} &
\colhead{Selection} &
\colhead{$T_{\rm BB}$} &
\colhead{Balmer break} &
\colhead{$f_{\rm BB}$}\\
\colhead{} &
\colhead{deg} &
\colhead{deg} &
\colhead{} &
\colhead{AB mag} &
\colhead{km\,s$^{-1}$} &
\colhead{erg\,s$^{-1}$} &
\colhead{AB mag} &
\colhead{} &
\colhead{} &
\colhead{K} &
\colhead{} &
\colhead{}\\
\colhead{(1)} &
\colhead{(2)} &
\colhead{(3)} &
\colhead{(4)} &
\colhead{(5)} &
\colhead{(6)} &
\colhead{(7)} &
\colhead{(8)} &
\colhead{(9)} &
\colhead{(10)} &
\colhead{(11)} &
\colhead{(12)} &
\colhead{(13)}
}
\startdata
93986 & 268.486879 & 65.187836 & 2.346 & 22.14 & \nodata & 42.43 & $-19.91\pm0.04$ & S-shape & 1 & $2510\pm10$ & $0.54\pm0.04$ & $0.18\pm0.01$ \\
154612 & 268.542899 & 65.246349 & 2.427 & 26.05 & \nodata & 41.44 & $-17.49\pm0.51$ & S-shape & 1 & $2910\pm180$ & $3.97\pm3.74$ & $0.34\pm0.10$ \\
102643 & 268.525941 & 65.227708 & 2.866 & 23.53 & \nodata & 42.76 & $-18.65\pm0.39$ & S-shape & 2 & $4010\pm40$ & $1.79\pm0.30$ & $0.57\pm0.02$ \\
92199 & 268.440149 & 65.197540 & 2.877 & 24.54 & $866\pm412$ & 42.20 & $-18.73\pm0.18$ & L-shape & 0 & $2500\pm30$ & $0.74\pm0.24$ & $0.10\pm0.01$ \\
109064 & 268.291743 & 65.192654 & 3.218 & 24.72 & $673\pm346$ & 42.20 & $-18.54\pm0.28$ & S-shape & 0 & $2970\pm130$ & $0.92\pm0.20$ & $0.22\pm0.04$ \\
31215 & 268.432888 & 65.147653 & 3.218 & 23.76 & $2290\pm112$ & 42.60 & $-18.08\pm0.63$ & S-shape & 0 & $3210\pm50$ & $1.08\pm0.41$ & $0.56\pm0.04$ \\
101522 & 268.505631 & 65.233142 & 3.472 & 23.34 & $2020\pm78$ & 42.71 & $-18.51\pm0.28$ & S-shape & 0 & $2930\pm40$ & $1.14\pm0.21$ & $0.50\pm0.03$ \\
90371 & 268.484841 & 65.207206 & 3.571 & 23.73 & $1082\pm112$ & 42.65 & $-17.78\pm0.56$ & S-shape & 1,2 & $4370\pm30$ & $3.14\pm0.68$ & $0.94\pm0.01$ \\
94366 & 268.502027 & 65.186014 & 3.661 & 25.75 & $1962\pm185$ & 42.14 & $-18.45\pm0.37$ & S-shape & 1 & $4220\pm270$ & $1.65\pm1.54$ & $0.60\pm0.08$ \\
150487 & 268.532516 & 65.267701 & 3.673 & 24.43 & $1293\pm106$ & 42.55 & $-18.87\pm0.16$ & S-shape & 1 & $3790\pm50$ & $0.93\pm0.23$ & $0.65\pm0.02$ \\
86363 & 268.392363 & 65.243382 & 3.829 & 24.09 & $1666\pm85$ & 42.60 & $-17.98\pm0.32$ & S-shape & 1 & $4430\pm40$ & $2.33\pm0.63$ & $0.92\pm0.01$ \\
93206 & 268.432919 & 65.191916 & 3.866 & 25.23 & $1628\pm214$ & 41.83 & $-17.51\pm0.53$ & S-shape & 1 & $4350\pm80$ & $8.65\pm8.46$ & $0.98\pm0.02$ \\
153244 & 268.394589 & 65.254180 & 3.882 & 24.69 & $1028\pm134$ & 42.60 & $-19.25\pm0.15$ & S-shape & 1 & $3160\pm140$ & $1.04\pm0.22$ & $0.41\pm0.06$ \\
45530 & 268.446213 & 65.134242 & 3.950 & 24.86 & $1404\pm248$ & 42.26 & $-18.20\pm0.45$ & S-shape & 2 & $4430\pm100$ & $2.12\pm0.91$ & $1.00\pm0.00$ \\
30608 & 268.416011 & 65.152857 & 3.972 & 25.24 & $1130\pm291$ & 41.97 & $-16.61\pm1.56$ & S-shape & 1 & $4130\pm190$ & $2.03\pm2.68$ & $0.87\pm0.08$ \\
112998 & 268.316214 & 65.169472 & 3.976 & 25.09 & $1450\pm177$ & 42.12 & $-17.66\pm0.61$ & S-shape & 1 & $2790\pm160$ & $0.93\pm0.29$ & $0.38\pm0.09$ \\
150447 & 268.426057 & 65.267825 & 4.047 & 25.41 & $1474\pm46$ & 42.61 & $-19.06\pm0.13$ & S-shape & 1 & $4430\pm200$ & $1.42\pm0.30$ & $0.45\pm0.06$ \\
89443 & 268.404027 & 65.212111 & 4.051 & 24.44 & $1706\pm95$ & 42.61 & $-18.71\pm0.32$ & S-shape & 2 & $3980\pm210$ & $1.14\pm0.44$ & $0.70\pm0.11$ \\
98249 & 268.441027 & 65.167628 & 4.397 & 25.07 & $1453\pm163$ & 42.22 & $-19.16\pm0.25$ & S-shape & 1 & $4480\pm150$ & $2.79\pm1.31$ & $0.73\pm0.03$ \\
44927 & 268.541892 & 65.138807 & 4.427 & 23.61 & $2595\pm0$ & 43.20 & $-18.69\pm0.25$ & V-shape & 1 & $2970\pm40$ & $1.60\pm0.46$ & $0.74\pm0.04$ \\
90510 & 268.445118 & 65.206495 & 4.513 & 25.99 & $458\pm263$ & 42.13 & $-18.40\pm0.31$ & L-shape & 0 & $3290\pm310$ & $2.47\pm2.19$ & $0.35\pm0.08$ \\
151449 & 268.522542 & 65.262644 & 4.524 & 23.34 & $2623\pm0$ & 43.51 & $-19.29\pm0.20$ & S-shape & 1 & $4700\pm40$ & $3.40\pm0.52$ & $0.81\pm0.04$ \\
80109 & 268.588853 & 65.173888 & 4.652 & 26.47 & \nodata & 41.64 & $-18.55\pm0.43$ & S-shape & 1 & $4020\pm590$ & $2.26\pm3.33$ & $0.30\pm0.20$ \\
98270 & 268.437673 & 65.167487 & 4.655 & 25.00 & $2655\pm21$ & 42.89 & $-18.38\pm0.42$ & S-shape & 1 & $4210\pm180$ & $2.10\pm0.82$ & $0.78\pm0.09$ \\
86487 & 268.381841 & 65.242666 & 5.006 & 25.70 & $1180\pm185$ & 42.16 & $-17.86\pm0.56$ & V-shape & 1 & $3430\pm270$ & $1.25\pm0.37$ & $0.71\pm0.12$ \\
96854 & 268.472997 & 65.174102 & 5.073 & 25.23 & $1055\pm249$ & 42.28 & $-18.37\pm0.80$ & L-shape & 1,2 & $3120\pm600$ & $1.89\pm0.78$ & $0.26\pm0.09$ \\
86558 & 268.408439 & 65.242195 & 5.100 & 23.47 & $2755\pm0$ & 43.32 & $-19.87\pm0.10$ & S-shape & 1,2,3 & $4640\pm50$ & $2.43\pm0.32$ & $0.86\pm0.02$ \\
91298 & 268.379572 & 65.202333 & 5.169 & 25.84 & $982\pm206$ & 41.93 & $-17.84\pm0.71$ & V-shape & 3 & $3350\pm290$ & $0.61\pm1.42$ & $0.71\pm0.14$ \\
110515 & 268.305972 & 65.183170 & 5.170 & 24.53 & $2011\pm92$ & 42.68 & $-18.54\pm0.38$ & S-shape & 1,2,3 & $4660\pm120$ & $1.34\pm0.55$ & $0.80\pm0.05$ \\
117491 & 268.303866 & 65.228111 & 5.264 & 25.09 & $1978\pm177$ & 42.31 & $-18.38\pm0.49$ & S-shape & 2,3 & $4390\pm120$ & $5.91\pm4.16$ & $0.99\pm0.06$ \\
96977 & 268.428335 & 65.173503 & 5.871 & 24.36 & $2430\pm117$ & 42.87 & $-19.15\pm0.35$ & V-shape & 1,3 & $2690\pm150$ & $2.45\pm1.51$ & $0.61\pm0.05$ \\
96784 & 268.568555 & 65.174341 & 5.873 & 23.77 & $2745\pm64$ & 43.16 & $-18.89\pm0.26$ & S-shape & 1,2,3 & $4330\pm60$ & $3.70\pm0.74$ & $0.83\pm0.02$ \\
93228 & 268.361906 & 65.191893 & 6.573 & 25.53 & $1264\pm123$ & 42.48 & $-19.68\pm0.18$ & V-shape & 1,3 & $3340\pm400$ & $1.59\pm0.70$ & $0.70\pm0.06$ \\
87572 & 268.497305 & 65.221068 & 6.700 & 25.50 & $1809\pm141$ & 42.77 & $-19.20\pm0.24$ & L-shape & 1 & $4550\pm910$ & $1.03\pm0.40$ & $0.65\pm0.16$ \\
96972 & 268.413441 & 65.173551 & 7.281 & 25.87 & $1772\pm611$ & \nodata & $-18.09\pm0.79$ & V-shape & 1,3 & $2900\pm340$ & $0.65\pm1.26$ & $0.85\pm0.06$ \\
92879 & 268.453402 & 65.194036 & 7.403 & 25.94 & $2799\pm489$ & \nodata & $-20.08\pm0.15$ & L-shape & 1,2,3 & $3140\pm400$ & $1.25\pm0.80$ & $0.49\pm0.09$ \\\enddata
\tablecomments{
Column (1): QDR ID.
Columns (2) and (3): J2000 right ascension and declination.
Column (4): spectroscopic redshift.
Column (5): F444W magnitude.
Column (6): FWHM corrected for instrumental broadening using the estimate from \citet{deGraaff_2024}. For sources at $z<7$, we list the broad H$\alpha$ FWHM; for sources at $z\geq7$, we list the broad H$\beta$ FWHM.
Column (7): broad H$\alpha$ luminosity.
Column (8): absolute UV magnitude, $M_{\rm UV}$.
Column (9): shape subclass, i.e., S-shape, V-shape, or L-shape.
Column (10): selection flag among the three photometric methods described in Section~\ref{sec:data}, where 1=V-color, 2=V-slope, and 3=Red-color.
Objects not recovered by any photometric method are marked as 0.
Column (11): best-fit blackbody temperature from the toy decomposition model.
Column (12): Balmer-break strength measured from the spectrum.
Column (13): blackbody fraction from the toy decomposition model.
}
\end{deluxetable*}
\end{rotatetable*}

We also experimented with a double-Gaussian model. However, given the low resolution of the PRISM spectra (\(R \sim 100\)--300), it is difficult to robustly decompose narrow and broad components. Only a few sources show clear evidence for an additional component. The limited resolution also prevents us from reliably modeling detailed line-profile features, such as absorption features or extended exponential wings, which have been discussed extensively in the recent literature \citep[e.g.,][]{LinXiaojing_2026,Torralba_2026,Matthee_2026}. We therefore note that the single-Gaussian approximation may overestimate the luminosity of the broad component and slightly underestimate its line width. Nevertheless, as shown in Figure~\ref{fig:fitting}, the single-Gaussian assumption provides a good description of the observed \Ha{} profiles, and the resulting systematic uncertainty does not significantly affect the following analysis.

For the \Hb+\OIII{} region, we fit the spectrum over a rest-frame window of 4650--5200\,\AA. In this complex, \Hb{} is modeled with a single Gaussian whose width is allowed to vary freely, while [O\,III]\(\lambda\lambda4959,5007\) are each modeled with a single Gaussian whose FWHM is fixed to the instrumental resolution. All components in the \Hb+\OIII{} complex are required to share the same velocity offset, and the [O\,III] \(\lambda4959:\lambda5007\) flux ratio is fixed to 1:3. The instrumental broadening is estimated using the wavelength-dependent PRISM spectral resolution curve from \citet{deGraaff_2024}, which provides a better representation of the point-source resolution than the nominal JDox curve \citep{JDox_2016}.

Figure~\ref{fig:fitting} presents the spectroscopic fitting results. Because the PRISM spectral resolution improves toward redder wavelengths, \Ha{} is generally better resolved and can be modeled more reliably than the \Hb+\OIII{} complex. As discussed above, the low resolution may produce artificial broad wings, which can drive the fit toward an extremely broad but weak additional component. We therefore adopt the simpler model with fewer degrees of freedom. Overall, the single-Gaussian model provides an adequate description of the line profiles and is sufficient to identify the broad-line nature of most LRDs.

In Table~\ref{tab:catalog}, we list the best-fit FWHM measurements, which are used to define our broad-line LRD sample. Among the full sample of 36 spectroscopic LRDs, 30 LRDs have resolved \Ha{} emission lines, while three LRDs at $z<2.87$ have both \Ha{} and \Hb{} unresolved in the PRISM spectra, as shown in Figure~\ref{fig:fitting}. In addition, ID-80109, with ${\rm F444W}=26.47$ mag, is too faint for robust line-profile modeling and is therefore also best fit as unresolved. For the remaining LRDs at $z<7$, we use the best-fit \Ha{} measurements for further analysis. For the two LRDs at $z>7$ without \Ha{} coverage, we use \Hb{} instead. We adopt a broad-line threshold of 600 km\,s$^{-1}$, following \citet{ZhuangMingyang_2026_NEXUS_EDR}. 31 well-modeled LRDs satisfy this broad-line criterion. Even if we adopt a stricter threshold of 1000 km\,s$^{-1}$ for broad line FWHM \citep[e.g.,][]{Harikane_2023,Greene_2024_Vshape_UNCOVER,Matthee_2024,LinXiaojing_2026}, about 90\% of the sample would be retained, further supporting the observation that most spectroscopically confirmed LRDs exhibit broad emission lines\citep{Hviding_2025}. Therefore, we simply adopt the 600~km\,s$^{-1}$ threshold and use the subsample of 31 broad-line LRDs for the analysis of SMBH properties in Section~\ref{sec:SMBH}.

\subsection{Physical Properties of the SMBH} \label{sec:SMBH}

We measure the total \Ha{} luminosity of the LRD sample from the best-fit models. To estimate the broad \Ha{} luminosity, we adopt a median broad-to-total \Ha{} ratio of 60\% \citep{Taylor_2025}. We therefore convert the total \Ha{} luminosity to broad \Ha{} luminosity ($L_{\rm H\alpha,broad}$) using this factor, and include an additional systematic uncertainty of 0.2 dex.

\begin{figure*}
\epsscale{1.14}
\plotone{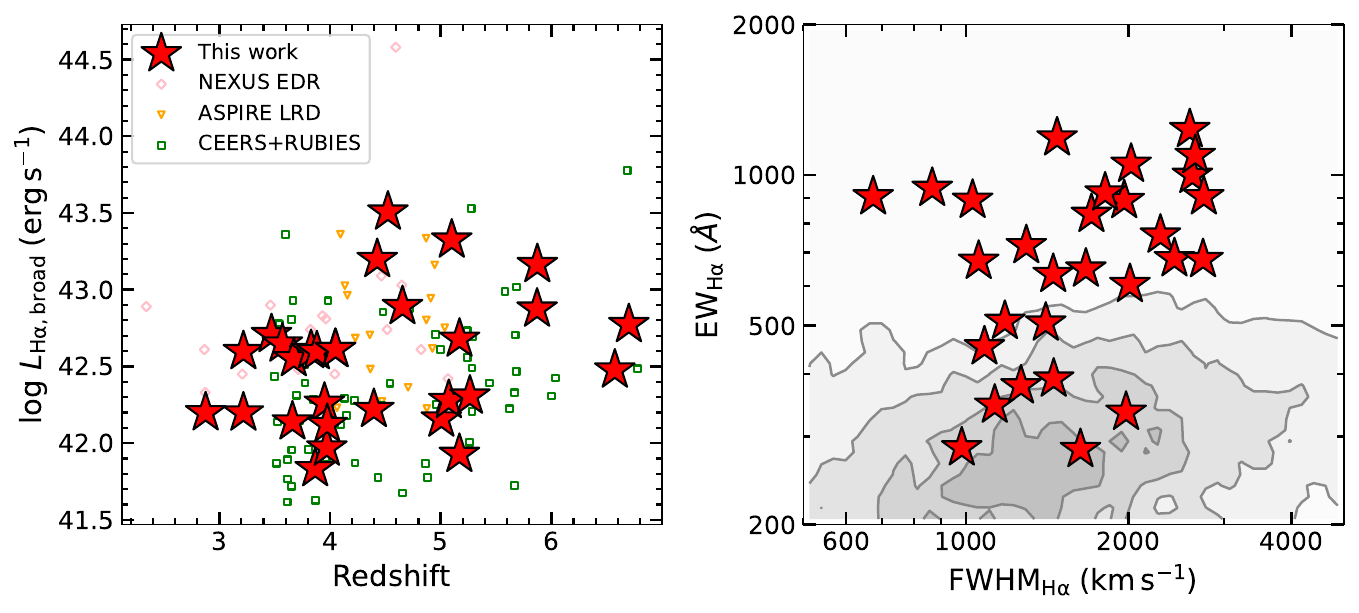}
\caption{
Left: Broad \Ha{} luminosity as a function of redshift for the LRDs in this work (filled red stars), compared with NEXUS EDR LRDs (pink diamonds; \citealt{ZhuangMingyang_2026_NEXUS_LRD}), ASPIRE LRDs (orange triangles; \citealt{LinXiaojing_2024}), and CEERS+RUBIES BLAGNs (green squares; \citealt{Taylor_2025}). Right: Rest-frame total \Ha{} equivalent width versus \Ha{} FWHM. Gray contours and shaded regions show the distribution of SDSS DR16 quasars at $z<0.6$.
}\label{fig:L_z_M}
\end{figure*}

In the left panel of Figure~\ref{fig:L_z_M}, we compare the redshift and $L_{\rm H\alpha,broad}$ of the NEXUS LRD sample with sources from other surveys, including broad-line AGNs (BLAGNs)/LRDs selected with NIRCam/WFSS spectroscopy from the NEXUS EDR \citep{ZhuangMingyang_2026_NEXUS_LRD} and ASPIRE surveys \citep{LinXiaojing_2024}, as well as BLAGNs selected with NIRSpec spectroscopy from the CEERS and RUBIES surveys \citep{Taylor_2025}. The NEXUS LRD sample in this work spans a wider redshift range ($2.3 < z < 7.5$) than any individual comparison sample. Its broad \Ha{} luminosities also cover a wide range, from $\sim 10^{42}$ to $10^{43.5},{\rm erg\,s^{-1}}$. In particular, this work reaches at least 0.5 mag deeper than the NEXUS EDR LRD sample based on NIRCam/WFSS spectra, enabling a deeper and more complete estimate of the LRD abundance at $3<z<5$, as discussed in Section~\ref{sec:abundance}.

We show the rest-frame total \Ha{} equivalent width (EW) versus FWHM distributions in the right panel of Figure~\ref{fig:L_z_M}. The instrumental correction is based on the wavelength-dependent resolution estimate from \citet{deGraaff_2024}. Compared to quasars at $z<0.6$ from the SDSS DR16Q catalog \citep{WuQiaoya_2022}, our high-redshift LRDs show a similar FWHM distribution but are systematically shifted toward higher EWs, suggesting that they may have a different physical origin from typical AGNs. We further discuss possible scenarios in Section~\ref{sec:decompose}.

Recent studies have suggested that the bolometric luminosities of LRDs may be nearly an order of magnitude lower than those inferred using standard prescriptions for type-1 AGNs \citep[e.g.,][]{Greene_2026_LRDLbol,Umeda_2026}. Even among normal AGNs, different samples with different physical properties can yield different average bolometric correction factors \citep{Stern_2012,ShenYue_2011_DR7Q,WuQiaoya_2022}. It is currently unclear whether or not these relations calibrated for low-redshift type-1 AGNs are applicable to high-redshift BLAGNs or LRDs, or which relation would be the most representative. We thus do not report bolometric luminosities as fiducial measurements in this work. Nevertheless, for qualitative number-density comparisons with previous studies in Section~\ref{sec:abundance}, we convert the literature bolometric luminosity to equivalent broad-\Ha{} luminosity using the empirical relation $L_{\rm bol}=130\times L_{\rm H\alpha,broad}$ from \citet{Stern_2012}, as adopted in other LRD studies.

The standard single-epoch virial formulae for black hole mass estimates are calibrated using local type-1 AGNs and may not be directly applicable to LRDs, whose continuum and broad-line emission mechanisms may differ substantially from those of typical unobscured AGNs. In particular, recent models suggest that dense ionized gas surrounding the black hole can broaden the observed Balmer-line profiles through electron scattering, causing virial BH masses inferred from the observed FWHMs to be overestimated by $\sim1$--2 dex \citep[e.g.,][]{Rusakov_2026}. Therefore, we choose not to report black hole masses estimated from traditional single-epoch virial relations. Nevertheless, as a sanity check, we have tested such measurements using the \Ha{}-based relation from \citet{Reines_2013} and the \Hb{}-based relation from \citet{Vestergaard_2006}. The resulting estimates place our LRDs in a similar region of the $L_{\rm H\alpha,broad}$--$M_{\rm BH}$ plane as BLAGNs and LRDs reported in NEXUS EDR and other JWST surveys. We defer a more detailed exploration of bolometric luminosity and blcak hole mass estimates to future work.

Obtaining reliable stellar masses for the host galaxies of LRDs is also extremely challenging. In most cases, the rest-frame optical emission is only marginally resolved, making the nuclear and host-galaxy components highly degenerate. In addition, the current imaging depth is insufficient to enable robust image decomposition for the majority of the sample. For these reasons, we provide only rough stellar-mass estimates in this work. We first estimate $M_{\rm UV}$ from the median flux density around 1500\,\AA. For sources whose spectra do not cover this wavelength range, we instead use the F090W photometry as a proxy to estimate $M_{\rm UV}$. 

We assume that the UV emission is entirely dominated by the host galaxy and estimate stellar masses using the $M_\ast$--$M_{\rm UV}$ relation from \citet{Song_2016}. These estimates should therefore be regarded as highly uncertain. Under this assumption, most LRDs have inferred stellar masses of $\sim10^{6}$--$10^{8},M_\odot$ and lie $\sim1$--$2$ dex above the local $M_{\rm BH}$--$M_\ast$ relation for classical bulges and elliptical galaxies \citep{Kormendy_Ho_2013}. This offset may further suggest that traditional single-epoch black hole mass scalings are not directly applicable to LRDs \citep[e.g.,][]{Rusakov_2026}.

\subsection{Abundance and Luminosity function} \label{sec:abundance}

We use Hierarchical Equal Area isoLatitude Pixelization (HEALPix; \citealt{Gorski_2005_HEALPix,Zonca_2019_Healpy}) to more accurately estimate the effective area of the current MSA footprint. We adopt an appropriate base resolution for the interior regions and refine the resolution near the footprint boundaries, yielding an effective area of about $66~{\rm arcmin}^2$. As mentioned in Sections~\ref{sec:NEXUS} and \ref{sec:comp}, the overall spectroscopic completeness for sources with ${\rm F444W}<26$ mag is about 73\%. To account for spectroscopic incompleteness more carefully, we estimate the completeness separately for targets with different MSA priorities, which ranges from 95\% to 61\%.

Taking the spectroscopic incompleteness into consideration, we estimate the binned broad-\Ha{} luminosity function, $\Phi$, of the NEXUS MSA LRD sample as a function of $\log L_{\rm H\alpha,broad}$ over $2.3<z<6.7$, using the 34 LRDs with available \Ha{} measurements. The results are listed in Table~\ref{tab:LF}. The statistical uncertainties are assumed to follow Poisson counting statistics, with small-number corrections estimated using Equation~(7) of \citet{Gehrels_1986}. This estimate is expected to be relatively robust over most of the luminosity and redshift range considered here. However, the faintest luminosity bin and the highest-redshift bin should be regarded as approximate lower limits, since our correction only accounts for spectroscopic incompleteness associated with MSA target priorities and does not include other sources of incompleteness that may have the largest impact at the faintest and highest-redshift ends.

\begin{figure*}
\epsscale{1.14}
\plotone{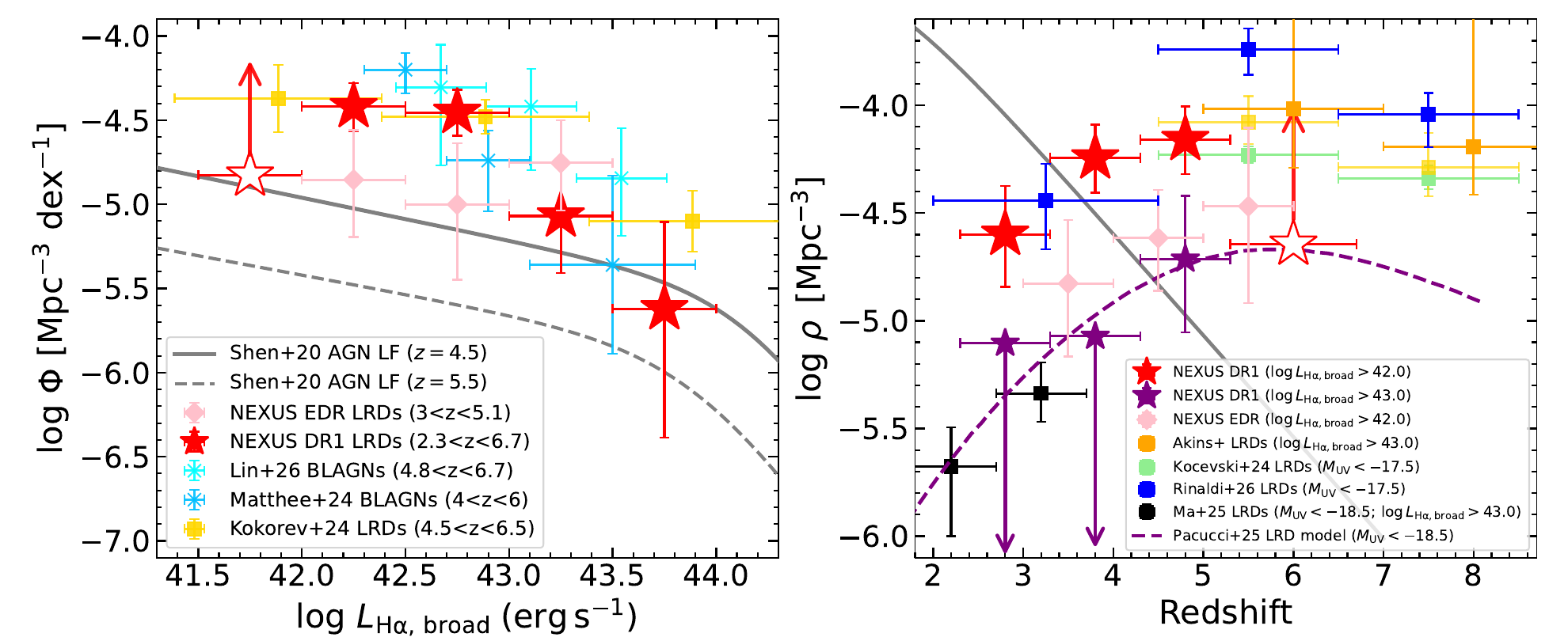}
\caption{
Left: Broad \Ha{} luminosity functions of LRDs at $z>2$. Red stars represent the NEXUS DR1 measurements over $2.3<z<6.7$, and pink diamonds show the NEXUS EDR measurements over $3<z<5.1$. Cyan crosses, blue crosses, and gold squares show literature results from \citet{LinXiaojing_2026}, \citet{Matthee_2024}, and \citet{Kokorev_2024_Vshape}, respectively. Gray curves show the extrapolated AGN bolometric luminosity functions from \citet{ShenXuejian_2020} at $z=4.5$ and 5.5, converted to broad \Ha{} luminosity using the same bolometric correction adopted in this work. 
Right: Accumulated comoving number density of LRDs as a function of redshift. Red stars show the DR1 measurements with $\log L_{\rm H\alpha,broad}>42.0$; the highest-redshift bin is shown as an open symbol with an upward arrow to indicate that it should be regarded as a lower limit owing to substantial incompleteness. Purple stars show the DR1 measurements with a brighter cut of $\log L_{\rm H\alpha,broad}>43.0$, used as a benchmark for comparison with literature. The downward arrow denotes the $1\sigma$ upper limit for a redshift bin with no detected sources. Pink diamonds show the EDR measurements. Orange, light-green, blue, and black squares represent literature measurements from \citet{Akins_2025_Cosmos_Red}, \citet{Kocevski_2025_Slope}, \citet{Rinaldi_2026}, and \citet{MaYilun_2025}, respectively. The purple dashed curve shows the theoretical LRD density evolution at $M_{\rm UV}<-18.5$ from \citet{Pacucci_2025}. Vertical error bars indicate $1\sigma$ uncertainties, while horizontal error bars indicate bin widths in luminosity (left) and redshift (right).
}
\label{fig:abundance}
\end{figure*}

The left panel of Figure~\ref{fig:abundance} shows our estimate of the binned broad \Ha{} luminosity function, with red stars denoting the NEXUS DR1 measurements. At the bright end, $\log L_{\rm H\alpha,broad}>42.5$, the inferred number densities are broadly consistent with literature results, suggesting that the completeness assumption is reasonably adequate in this luminosity range. In contrast, at the faintest end, $\log L_{\rm H\alpha,broad}<42$, the inferred number density is significantly lower, indicating that a more careful incompleteness analysis is required. For example, future work should quantify how spectroscopic depth, including a practical limit of $\sim 26$ mag, affects the identification of faint LRDs as a function of redshift and $\log L_{\rm H\alpha,broad}$.

We also note that, at $\log L_{\rm H\alpha,broad}<42.5$, the number densities inferred from the NEXUS EDR LRDs are lower than the estimates from this work using MSA spectra, and other literature estimates. This discrepancy likely reflects the shallower depth and lower completeness of the NEXUS EDR WFSS LRD sample in this luminosity range. As pointed out in \citet{ZhuangMingyang_2026_NEXUS_LRD}, their density estimate does not account for sensitivity limits or the detectability of broad emission lines. We visually inspect the WFSS spectra of the DR1 LRD sample and confirm that some objects show only a single emission line, or even no clear line, within the wavelength coverage. This makes it difficult to establish their AGN nature or obtain secure redshifts. In addition, contamination from bright sources in the WFSS data may further reduce the number of identifiable LRDs. The EDR estimate should therefore be regarded as a lower limit on the LRD number density.

\begin{deluxetable}{Cccc}
\caption{Broad H$\alpha$ Luminosity Function at $2.3<z<6.7$}
\label{tab:LF}
\tablehead{
\colhead{$\log L_{\rm H\alpha,broad}$} & \colhead{$\log L_{\rm H\alpha,broad}$ Range} & \colhead{$N$} & \colhead{$\Phi$}
\\
\colhead{$({\rm erg\,s^{-1}})$} & \colhead{$({\rm erg\,s^{-1}})$} & \colhead{} & \colhead{$(10^{-6}\,{\rm Mpc}^{-3}\,{\rm dex}^{-1})$}
}
\startdata
$41.700$ & $41.40<\log L<42.00$ & $5$ & $14.43_{-6.22}^{+9.74}$ \\
$42.250$ & $42.00<\log L<42.50$ & $12$ & $38.09_{-10.83}^{+14.46}$ \\
$42.750$ & $42.50<\log L<43.00$ & $13$ & $35.15_{-9.61}^{+12.69}$ \\
$43.250$ & $43.00<\log L<43.50$ & $3$ & $8.52_{-4.62}^{+8.26}$ \\
$43.750$ & $43.50<\log L<44.00$ & $1$ & $2.40_{-1.98}^{+5.48}$ \\
\enddata
\tablecomments{The number densities have been corrected for the spectroscopic incompleteness of the current six epochs of MSA observations. Uncertainties are corrected Poisson errors following \citet{Gehrels_1986}. }
\end{deluxetable}

To compare the LRD luminosity function with those of normal quasars/AGNs, we convert the quasar bolometric luminosity functions from \citet{ShenXuejian_2020} to broad \Ha{} luminosity functions using the same bolometric correction adopted in this work. We find that, at $\log L_{\rm H\alpha,broad}<43$, the number density of LRDs at $z\sim4.5$ is about 0.5 dex higher than the extrapolation of the quasar luminosity function. This result is consistent with previous studies \citep{Greene_2024_Vshape_UNCOVER,Matthee_2024,LinXiaojing_2026}, and further supports the prevalence of a population of faint broad-line AGNs and LRDs at high redshift that are missed by previous Type 1 quasar/AGN surveys around these redshifts. This overabundance of LRDs over normal AGNs is more prominent at even higher redshifts (Fig.~\ref{fig:abundance}) given the opposite evolutionary trends of LRD and AGN number densities towards higher redshift.  

Table~\ref{tab:density} and the right panel of Figure~\ref{fig:abundance} show the accumulated comoving number density, $\rho$, as a function of redshift. We compute the LRD number densities in several redshift bins after correcting for spectroscopic incompleteness, and compare our results with literature measurements for AGN and LRD samples. We attempt to homogenize the literature measurements to the same $\log L_{\rm H\alpha,broad}$ or $M_{\rm UV}$ threshold whenever possible. For the NEXUS EDR LRD sample from \citet{ZhuangMingyang_2026_NEXUS_LRD}, we estimate the abundance using a cut of $\log L_{\rm H\alpha,broad}>42$. For the LRDs from \citet{Akins2024_LRD}, we adopt a cut of $\log L_{\rm H\alpha,broad}>43$, corresponding to their reported value at $\log L_{\rm bol}>45$. For the LRDs from \citet{Kokorev_2024_Vshape} and \citet{Rinaldi_2026}, we use a common threshold of $M_{\rm UV}<-17.5$. We also show the theoretical prediction from \citet{Pacucci_2025} for $M_{\rm UV}<-18.5$. Finally, we include the $z<4$ estimates from \citet{MaYilun_2025}, using a threshold of $M_{\rm UV}<-18.5$, which is approximately comparable to their reported cut of $M_{5500}<-20.5$. The only spectroscopically confirmed LRD in their sample has $\log L_{\rm H\alpha,broad}=43.25$. Although there is no strong correlation between $\log L_{\rm H\alpha,broad}$ and $M_{\rm UV}$, for the purpose of comparison we approximately treat $\log L_{\rm H\alpha,broad}=43$ as corresponding to $M_{\rm UV}=-18.5$, and $\log L_{\rm H\alpha,broad}=42$ as corresponding to $M_{\rm UV}=-17.5$. We emphasize that this is only an illustrative approximation, since different studies may have different photometric and spectroscopic depths, as well as different methods for deriving $M_{\rm UV}$ and $\log L_{\rm H\alpha,broad}$.

\begin{deluxetable}{Cccc}
\tablecaption{\centerline{Accumulated Comoving Number Density}
\centerline{at $\log L_{\rm H\alpha,broad}>42$}\label{tab:density}}
\tablehead{
\colhead{Redshift} & \colhead{Redshift Range} & \colhead{$N$} & \colhead{$n\ (10^{-6}\,{\rm Mpc}^{-3})$}
}
\startdata
$2.80$ & $2.3<z<3.3$ & $5$ & $25.20_{-10.86}^{+17.01}$ \\
$3.80$ & $3.3<z<4.3$ & $10$ & $57.02_{-17.70}^{+24.30}$ \\
$4.80$ & $4.3<z<5.3$ & $10$ & $69.43_{-21.55}^{+29.59}$ \\
$6.00$ & $5.3<z<6.7$ & $4$ & $22.68_{-10.82}^{+17.89}$ \\
\enddata
\end{deluxetable}

Our measurements are broadly consistent with other JWST-based estimates at thresholds roughly comparable to $\log L_{\rm H\alpha,broad}>42$, although they are systematically higher than the EDR estimate by about $0.5$ dex. As shown in the left panel, while the nominal luminosity limit of the EDR LRD sample is also $\log L_{\rm H\alpha,broad}=42$, the more realistic limit for high completeness is likely closer to $\log L_{\rm H\alpha,broad}=43$. Under this brighter cut, the EDR estimates are more consistent with those from \citet{Pacucci_2025} and \citet{MaYilun_2025}. We also estimate the abundance of our sample using a brighter cut of $\log L_{\rm H\alpha,broad}>43$. Even though our sample does not contain LRDs with $\log L_{\rm H\alpha,broad}>43$ at $z<4$ due to the small volume probed by NEXUS, the measurement at $z\sim5$ shown by the purple star remains consistent with the literature results. This suggests that the apparent discrepancy between the estimates from \citet{MaYilun_2025} and our measurements can be largely alleviated once the different luminosity limits are taken into account. 

Overall, our estimates of the LRD abundance are reasonable and provide the first spectroscopic constraint on the LRD abundance down to $\log L_{\rm H\alpha,broad}>42$ at $2.5<z<5$ with high completeness. We also plot the accumulative AGN number density for $L_{\rm bol}>10^{44}\,\mathrm{erg\,s^{-1}}$ using the extrapolated luminosity function from \citet{ShenXuejian_2020} in the right panel of Figure~\ref{fig:abundance}. Our measurements indicate that the number density of high-redshift LRDs is significantly higher than that of normal AGNs by about $0.5$--1 dex, consistent with previous findings \citep[e.g.,][]{LinXiaojing_2026}. Meanwhile, the abundance of LRDs appears to decrease toward lower redshift. Future wide-area studies with the \textit{Roman Space Telescope}, targeting AGNs and LRDs at cosmic noon, will be essential for further studying the evolution of the AGN and LRD populations toward lower redshifts and the faint end.

\subsection{Spatial Clustering}
Figure~\ref{fig:LRD_spatial} shows the spatial distribution of our spectroscopically confirmed LRDs (red filled circles) within the central NEXUS footprint. To quantify the clustering properties of this sparse spectroscopic LRD sample, we cross-correlate it with the much larger NEXUS photometric galaxy sample. Cross-correlation with a photometric sample also has the added advantage that the spectroscopic sample does not have to be homogeneously selected from the survey footprint. We restrict the photometric galaxy sample to sources with $22<m_{\rm F444W}<26$ and $2.3<z_{\rm phot}<6.7$, yielding 2166 galaxies in total. The redshift range is chosen to match that used for the abundance measurements in Section~\ref{sec:abundance}, ensuring consistency. For the LRD sample, we apply the same redshift cut to enable a meaningful cross-correlation measurement, leaving 34 objects in the final sample.

\begin{figure}
\epsscale{1.15}
\plotone{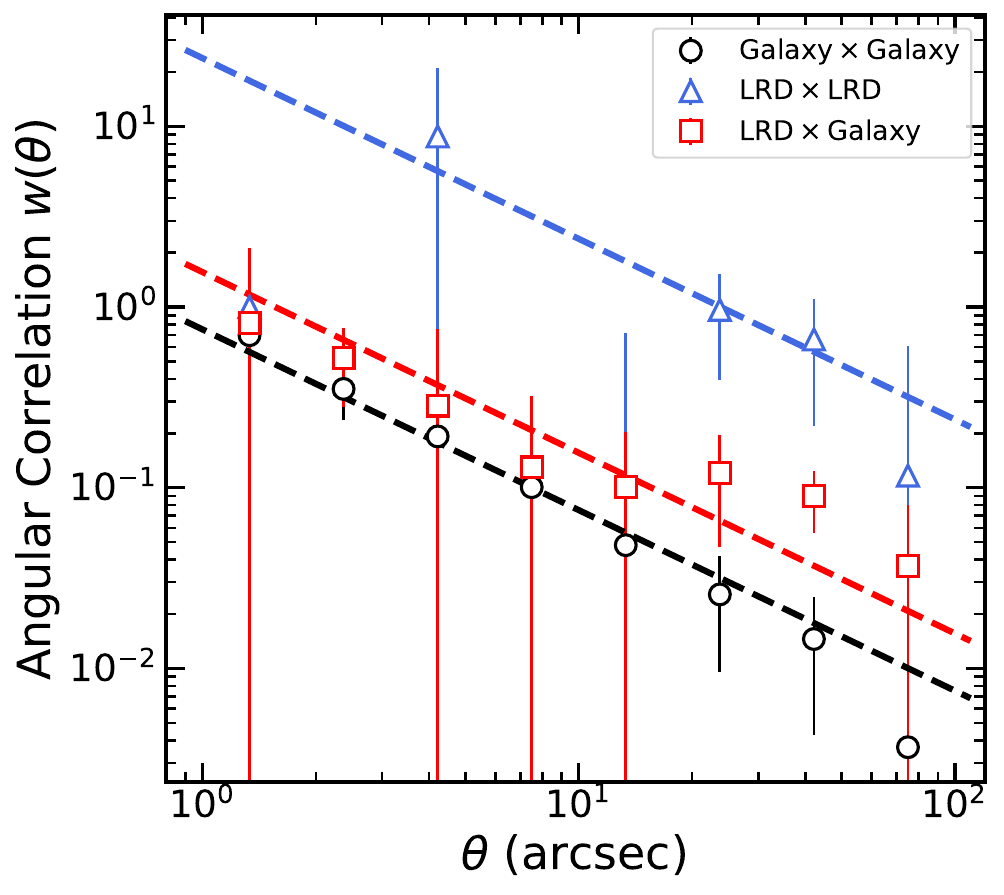}
\caption{
Angular correlation functions for different samples, measured over $1\lesssim \theta \lesssim 100$\arcsec. The best-fit power-law models are shown as dashed lines and summarized in Table~\ref{tab:clustering}. Uncertainties are estimated using 9 jackknife resampling regions.
}\label{fig:clustering}
\end{figure}

We measure the angular cross-correlation function (CCF) between the LRD sample and the photometric galaxy sample, and present the results in Figure~\ref{fig:clustering}. For comparison, we also show the auto-correlation functions (ACF) of the galaxy sample and the LRD sample as black open circles and blue triangles, respectively --  although we caution on the ACF of the spectroscopic LRD sample, since it may suffer from selection incompleteness that could create artificial spatial structures in the ACF. For the ACF, we use the Landy--Szalay estimator \citep{Landy_1993}. For the cross-correlation, we adopt the simple estimator QG/QR-1, where QG and QR are the normalized numbers of LRD--galaxy and LRD--random pairs in each angular separation bin. The uncertainties of both the ACF and CCF measurements are estimated using standard jackknife resampling.

Following \citet{ZhuangMingyang_2026_NEXUS_LRD}, we model the measured small-scale clustering as a power-law extrapolation of the large-scale two-point correlation function, $\omega(\theta)=(\theta/\theta_0)^{-\beta}=A_0\theta^{-\beta}$, where $\theta$ is the angular separation in units of radians. To reduce edge effects from the survey boundaries, we fit only the angular bins with $\theta<100\arcsec$. Given the large uncertainties due to the small sample size, we fix the power-law slope to $\beta=1$ for all samples, following \citet{ShenYue_2007}. We have also tested $\beta=0.8$, as adopted by \citet{ZhuangMingyang_2026_NEXUS_LRD}, and confirmed that this choice does not affect our conclusions. Assuming a real-space correlation function of the form $\xi(r)=(r/r_0)^{-\gamma}$ with $\gamma\equiv\beta+1=2$, we derive the real-space correlation length $r_0$ and the linear bias $b$ for both the ACF and CCF samples. Further details of the assumptions and calculations are provided in \citet{ZhuangMingyang_2026_NEXUS_LRD}.

\begin{table}
\caption{Clustering Measurements}\label{tab:clustering}
\centering
\begin{tabular}{llll}
\hline\hline
Sample & $\theta_0$ ($\beta=1$) & $r_0$ ($\gamma=2$) & linear bias \\
 & & $h^{-1}{\rm cMpc}$ & \\
\hline
Galaxy×Galaxy & $0.75_{-0.10}^{+0.10}$ & $2.69_{-0.19}^{+0.18}$ & $1.39_{-0.10}^{+0.09}$ \\
LRD×LRD & $23.85_{-10.25}^{+10.25}$ & $12.89_{-3.16}^{+2.52}$ & $8.20_{-2.01}^{+1.60}$ \\
LRD×Galaxy & $1.56_{-0.42}^{+0.42}$ & $5.37_{-0.79}^{+0.68}$ & $3.09_{-0.45}^{+0.39}$ \\

\hline
\hline\\
\end{tabular}
\end{table}

Table~\ref{tab:clustering} summarizes the clustering measurements for the different samples. We detect a significant clustering signal for the LRD sample over angular scales of $1\arcsec$--$100\arcsec$, which are well probed by our survey area, although the strength of the CCF is somewhat weaker than that reported by \citet{ZhuangMingyang_2026_NEXUS_LRD}. The linear bias inferred from the direct LRD ACF measurement is $b_{\rm LRD}=8.20_{-2.01}^{+1.6}$, consistent with the value of $b_{\rm LRD}=6.87_{-2.18}^{+2.54}$ obtained by assuming $b_{\rm LRD\times gal}^2 \sim b_{\rm LRD} b_{\rm gal}$, where $b_{\rm LRD}$ and $b_{\rm gal}$ are the linear biases of the LRD and galaxy samples, respectively. Our measurement is lower than the value of $b=10.15_{-4.54}^{+4.82}$ reported by \citet{ZhuangMingyang_2026_NEXUS_LRD}, but remains consistent within $1\sigma$. The linear bias value in this work roughly corresponds to typical halo masses of a few $\times 10^{11}\,h^{-1}M_\odot$, which is broadly consistent with recent clustering measurements of JWST BLAGNs at $z\sim4-6$ \citep[e.g.,][]{Arita_2025,LinXiaojing_2026_halo}. Improved sample statistics of LRDs from future NEXUS data will refine these clustering measurements.

\section{Discussion}   \label{sec:discuss}

\subsection{Composite LRD Spectra}
\begin{figure*}
\epsscale{1.15}
\plotone{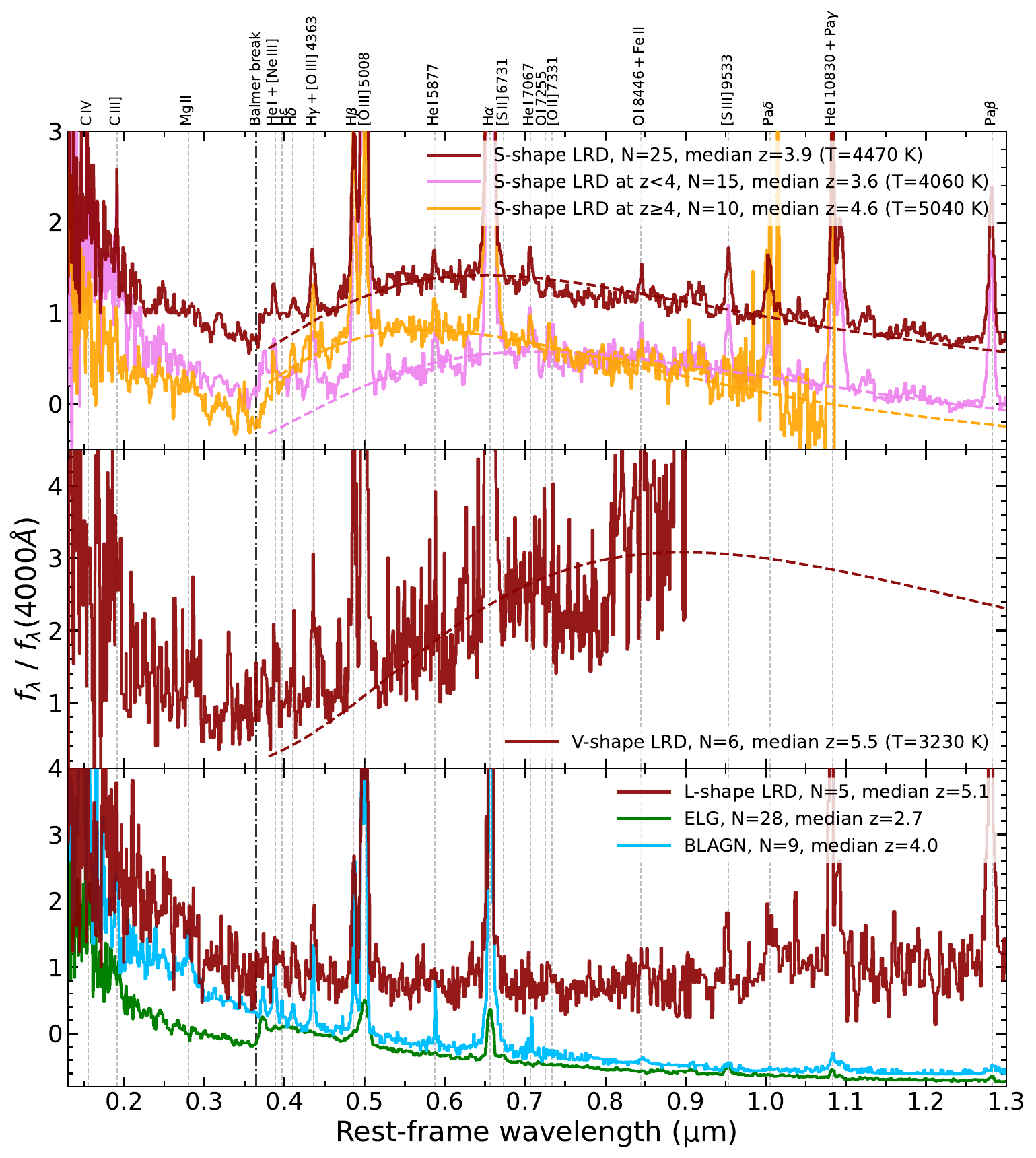}
\caption{
LRD composite spectra of the LRD and comparison samples in NEXUS, normalized by $f_\lambda$ at 4000\,\AA, smoothed, and shifted vertically for clarity. The number of objects included in each stack is listed in the corresponding label. The top panel shows the overall S-shape LRD composite (red) together with two redshift subsamples, $z<4$ (purple) and $z\geq4$ (orange). The middle panel shows the V-shape LRD composite. The bottom panel shows the L-shape LRD composite together with comparison composites for ELGs (green) and BLAGNs (blue) in NEXUS. Vertical gray dashed lines mark the wavelengths of major emission lines, and the black dot-dashed line marks the Balmer break at 3646\,\AA. For the S-shape and V-shape LRD composites, we additionally fit a blackbody model to the continuum redward of the Balmer break after masking the major emission-line regions. The corresponding best-fit blackbody continua are shown as dashed curves in the same color as the composite spectra, and the fitted temperatures are listed in the labels. The diverse spectral shapes of LRDs in rest-optical at different redshifts can be effectively explained by a simple blackbody continuum with a single characteristic temperature.
}\label{fig:composite}
\end{figure*}

As discussed in Section~\ref{sec:spec_LRD} and summarized in Table~\ref{tab:catalog}, we visually divide our spectroscopic LRD sample into three subclasses based on their spectral shapes: S-shape, V-shape, and L-shape. To further illustrate the spectral diversity of LRDs and examine the differences among these subclasses, we construct stacked spectra for each group and compare their composite spectra in Figure~\ref{fig:composite}. {Our SED-shape definition strongly depends on the rest-wavelength coverage by the MSA spectra, hence source redshift. So interpreting these shapes with physical origins is somewhat difficult. In the following discussion, we fit the SEDs with simple models, while keeping in mind the effects of the limited wavelength coverage when interpreting the results.}


We normalize the spectra by the flux density at 4000\,\AA\ and resample them onto a common rest-frame wavelength grid using \texttt{SpectRes} \citep{Carnall_2017_Spectres}. The top panel of Figure~\ref{fig:composite} shows the stack of the S-shape LRDs, together with two subsamples divided at $z=4$. The S-shape subclass represents the most typical LRD spectral energy distribution, characterized by a clear UV upturn, a strong Balmer break, and a continuum that gradually declines toward the rest-frame infrared, producing the characteristic laid-down S-shape. The middle panel shows the V-shape LRD stack, which also exhibits a significant UV upturn, but the Balmer breaks are somewhat less pronounced than those of the S-shape LRDs on average. V-shape LRDs may be truncated S-shape LRDs, due to incomplete rest-frame wavelength coverage at these high redshifts. Therefore, the shape classifications used here primarily reflect observational differences in the available spectra, rather than distinct physical origins.

For both the S-shape and V-shape LRDs, we fit a simple blackbody model to the continuum redward of the Balmer break after masking the major emission-line regions, following the BH* interpretation \citep{Naidu_2025,deGraaff_2025}. The median temperature of the S-shape LRDs is around 4500\,K, apparently higher than that of the V-shape LRDs. However, this difference is driven by the phenomenological classification rather than by an intrinsic physical difference. To illustrate this point, we divide the S-shape LRDs into two redshift bins. The two redshift subsamples broadly preserve the same overall S-shape, providing a useful sanity check on our visual classification. Nevertheless, they show different rest-optical slopes and therefore different median blackbody temperatures: the $z<4$ S-shape LRDs have lower fitted temperatures, while the $z>4$ S-shape LRDs have higher fitted temperatures. This apparent redshift evolution in average effective temperature is artificial: a low-temperature S-shape LRD at high redshift would be classified as a V-shape LRD by definition, because the truncation at rest-IR wavelength would prevent us from observing the redder rest-frame continuum to classify it as an S-shape. In other words, high-temperature LRDs are more likely to be classified as S-shape, removing such objects from the V-shape class. Consequently, as shown in the middle panel, the median fitted temperature of the V-shape LRDs is indeed low.


In summary, we do not interpret the difference between the S-shape and V-shape composites as evidence for any intrinsic difference or redshift evolution in the effective temperature distribution. A more detailed analysis of the distributions of temperature, redshift, and SED shape is presented in Section~\ref{sec:decompose} and in Figure~\ref{fig:decompose}.

Finally, the L-shape LRDs do not show an obvious Balmer break and therefore represent a more tentative subclass of LRDs. Nevertheless, we note that similar sources have been reported in the literature as LRDs \citep[e.g.,][]{Hviding_2025,deGraaff_2025}, and such objects likely lie near the boundaries of current photometric selection criteria. These L-shape LRDs can be easily confused with some ELGs and BLAGNs, so we compare the L-shape LRD stack with stacks of possible contaminants in NEXUS to highlight the differences. As shown in the bottom panel of Figure~\ref{fig:composite}, all of these objects exhibit blue UV continua, while BLAGNs generally show stronger UV emission lines, such as \MgII{} and \CIV{}, than the L-shape LRDs and ELGs. In addition, ELGs show a stronger Balmer break than the L-shape LRDs. In the rest-optical region, the L-shape continuum is redder than those of ELGs and BLAGNs. Continuum emission from a host galaxy may contribute to the rest-optical SED of L-shape LRDs, but the compact morphology of these LRDs limits this possibility, and the weak UV emission lines of L-shape LRDs distinguish them from normal AGNs. Nevertheless, for very faint objects, such as ID-92879 with $m_{\mathrm{F444W}}=25.94$, distinguishing these different scenarios could be challenging. Overall, we retain the L-shape subclass in our analysis. Because there are only five L-shape LRDs, they do not affect the main conclusions of this work.

\subsection{BH* model decomposition}\label{sec:decompose}

In this subsection, we perform a simple spectral decomposition that additionally includes the host-galaxy contribution in the rest-frame UV, within the BH* framework \citep[e.g.,][]{Naidu_2025,Sun_2026_BH*}, in order to further characterize the spectral diversity of the sample.

For the host galaxy emission, we construct six ELG templates by stacking NEXUS ELGs at $z>2$ in six bins of UV color, defined by $\mathrm{F115W}-\mathrm{F150W}$. These ELG templates are used as possible host-galaxy components for the LRDs. In the fitting procedure, we test each ELG template in combination with a blackbody component and select the combination that provides the best fit. The fit is performed in the rest frame after masking the major emission-line regions. We define the BH* fraction as the fraction of blackbody flux at 5500\,\AA\ continuum.

\begin{figure*}
\epsscale{1.15}
\plotone{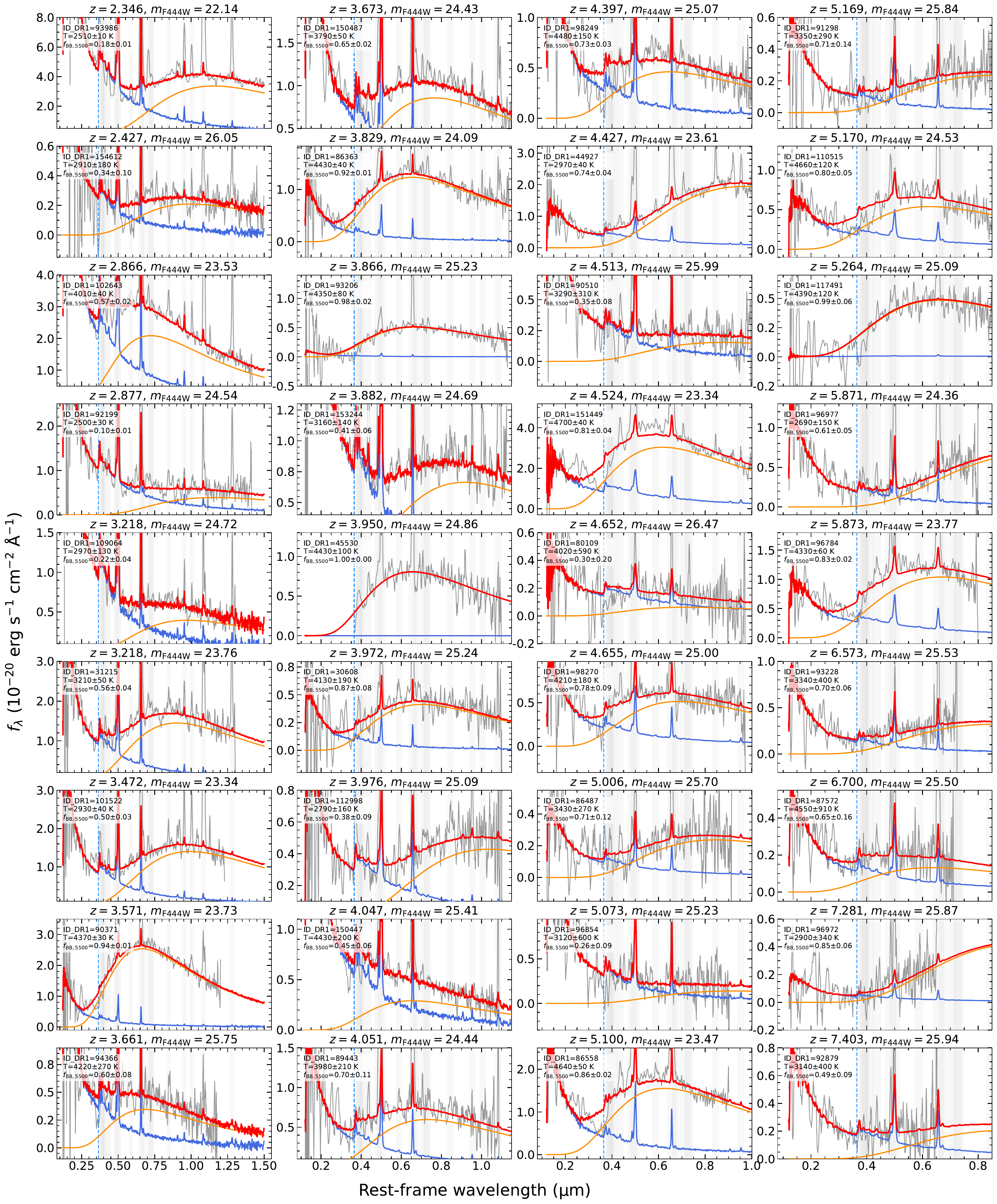}
\caption{
The host-galaxy plus blackbody continuum decomposition for each LRD, shown in the same format as Figure~\ref{fig:Spec_gallery}. The red curve shows the best-fit total model, while the blue and orange curves show the host-galaxy and blackbody components, respectively. Light gray shaded regions mark the emission-line windows excluded from the continuum fitting. The text in each panel lists the best-fit blackbody temperature and the blackbody flux fraction at 5500\,\AA.
}\label{fig:decompose}
\end{figure*}

Figure~\ref{fig:decompose} shows the best-fit decomposition for each LRD. Given the simplicity of the toy model, its performance is encouraging, as it successfully reproduces much of the observed spectral diversity of the sample. For the most typical S-shape LRDs, such as ID=90371, 86363, ID=151449, and ID=117491, the blackbody fraction is close to unity, indicating that these sources are nearly naked BH* systems, similar to literature examples such as \textit{MoM-BH*-1} \citep{Naidu_2025} and the Cliff \citep{deGraaff_2025_Cliff}.

For some lower-redshift S-shape LRDs, such as ID=93986, the apparent break occurs at wavelengths longer than the Balmer break. This behavior can be explained by a combination of stronger host-galaxy emission and a weaker, lower-temperature blackbody component. As the sources become fainter and the BH* fraction decreases further, the spectra can resemble those of the L-shape LRDs, such as ID=92199 and ID=96854. In these objects, the blackbody temperature is not well constrained, whereas the large host-galaxy contribution appears to be more robust.

\begin{figure*}
\epsscale{1.15}
\plotone{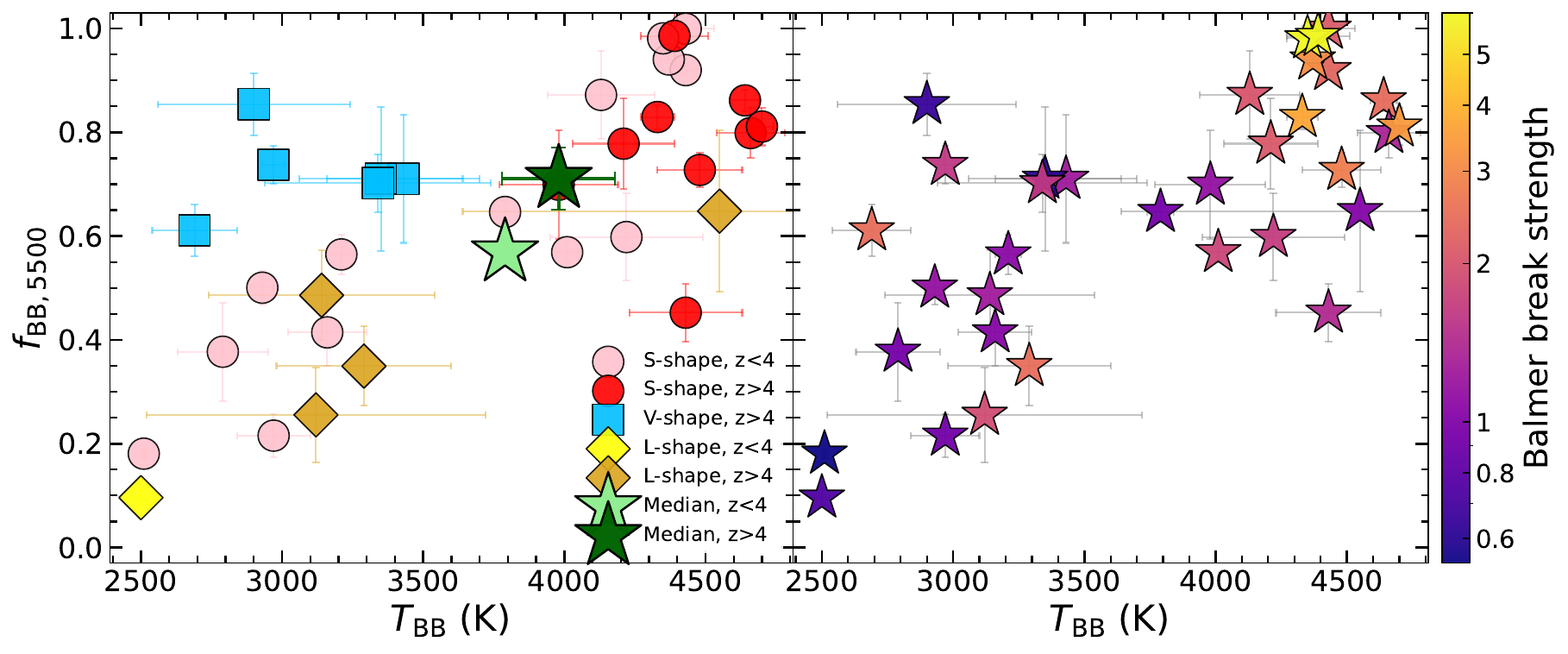}
\caption{
Left: Blackbody temperature versus blackbody flux fraction at 5500\,\AA\ for the spectroscopic LRD sample. Different symbol shapes denote the three spectral subclasses, S-shape, V-shape, and L-shape, while different colors indicate the two redshift intervals, $z<4$ and $z>4$. The light-green and dark-green stars show the median values for the $z<4$ and $z>4$ subsamples, respectively, with error bars indicating the median measurement uncertainties in each bin. Right: The same parameter space, but with all LRDs shown as stars and color-coded by Balmer break strength, which highlights how the blackbody temperature and BH* fraction relate to the observed Balmer-break strength. 
}\label{fig:BB_T}
\end{figure*}

In the left panel of Figure~\ref{fig:BB_T}, we show the distribution of blackbody temperature and BH* fraction, where the latter is quantified by the blackbody flux fraction at 5500\,\AA, for different spectral subclasses and redshift ranges. {This distribution provides a more direct illustration than Figure~\ref{fig:composite} for explaining the apparent redshift and temperature differences between S-shape and V-shape LRDs. Light-colored symbols represent low-redshift LRDs, while dark-colored symbols represent high-redshift LRDs. At low redshift, the sample contains only S-shape and L-shape LRDs, shown in pink and yellow, respectively, and their fitted temperatures span a broad range from $\sim2500$ to 5000\,K. There are no V-shape LRDs in this redshift range, because the available wavelength coverage is sufficient to reveal the full S-shape continuum. At $z>4$, high-temperature LRDs can still be classified as S-shape objects, shown as red points, whereas low-temperature LRDs can only be classified as either V-shape objects, shown in blue, or L-shape objects, shown in gold, depending on the relative host-galaxy contribution.}

To further test for possible redshift evolution, we also show the median results for LRDs at $z<4$ and $z>4$, shown in light green and dark green, respectively. Within the uncertainties, we find no significant difference in fitted temperature between the two redshift bins. The fitted BH* fraction differs by about 15\%, but this difference is most likely driven by observational and measurement biases. For example, at higher redshift, LRDs with stronger host-galaxy contributions may be harder to select and to constrain because of the more limited rest-frame wavelength coverage. Therefore, apart from the abundance evolution discussed above, we do not find clear evidence for redshift evolution in the fitted continuum properties of LRDs.

\begin{figure*}
\epsscale{1.15}
\plotone{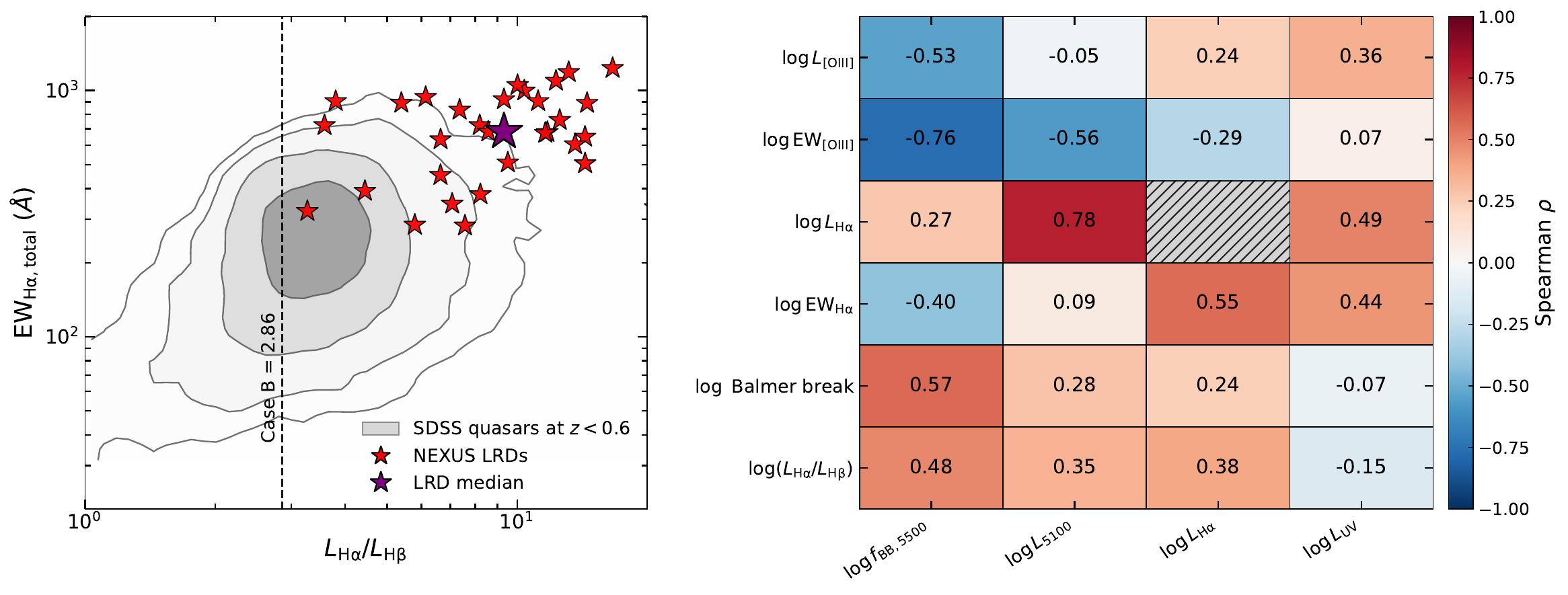}
\caption{
Left: Rest-frame total \Ha{} equivalent width versus Balmer decrement, $L_{\rm H\alpha}/L_{\rm H\beta}$. Gray contours show the counts-density distribution of SDSS DR16 quasars at $z<0.6$, red stars show the NEXUS LRDs, and the purple star marks the LRD median. The black dashed line indicates the Case B value of 2.86. Right: Spearman rank correlation matrix for LRD emission-line and continuum properties, restricted to sources with $M_{\rm UV}<-17.5$. Rows correspond to line luminosities, equivalent widths, Balmer-break strength, and Balmer decrement, while columns correspond to $\log f_{\rm BB,5500}$, $\log L_{5100}$, $\log L_{\rm H\alpha}$, and $\log L_{\rm UV}$. The color scale indicates the Spearman coefficient $\rho$.
}\label{fig:EW_BS}
\end{figure*}

In the right panel of Figure~\ref{fig:BB_T}, we show the same distribution, but color-coded by the Balmer-break strength. The Balmer-break strength is measured from the PRISM spectra as the ratio of the mean $f_\nu$ values in the rest-frame wavelength intervals [0.362, 0.372]\,\micron\ and [0.400, 0.410]\,\micron, following the definition commonly adopted in the literature \citep[e.g.,][]{deGraaff_2025}. For a few special cases, we slightly enlarge the measurement windows to ensure a more robust estimate. The Balmer-break strength can be regarded as a proxy for the overall continuum shape of LRDs. As shown in Figure~\ref{fig:BB_T}, it correlates well with both the fitted BH* fraction and blackbody temperature: LRDs with stronger Balmer breaks tend to have higher fitted blackbody temperatures and larger BH* fractions, corresponding to weaker host-galaxy contributions.

In addition, we estimate the blackbody luminosities from the best-fit temperatures. We find that the inferred blackbody luminosities are systematically lower than the bolometric luminosities, with a median offset of 0.8 dex. This is similar to the 0.9 dex offset reported by \citet{deGraaff_2025}, and supports the argument discussed in Section~\ref{sec:SMBH} that applying traditional bolometric corrections calibrated for normal AGNs may overestimate the bolometric luminosities of LRDs \citep{Greene_2026_LRDLbol,Umeda_2026}.

We also present additional evidence for the BH* scenario in Figure~\ref{fig:EW_BS}. In the left panel, we compare the exceptionally large total \Ha{} equivalent widths and Balmer decrements of the LRDs with a compilation of SDSS quasars at $z\simeq0$--0.6 \citep{WuQiaoya_2022}. The median value of the LRD sample, shown as the purple star, lies in the outskirt of the local quasar distribution, suggesting that the mechanisms powering the Balmer lines in LRDs may differ from those in typical quasars/AGNs.

In the right panel of Figure~\ref{fig:EW_BS}, we show a correlation matrix for different LRD properties, providing additional clues for the BH* scenario. The strongest correlation is between the \Ha{} luminosity and the continuum luminosity, $\log L_{5100}$, both estimated from the MSA spectra. In contrast, there is little correlation between ${\rm EW}_{\rm H\alpha}$ and $\log L_{5100}$. This suggests that the \Ha{} emission and the rest-frame optical continuum emission are likely linked to the same energy source, identified here as the BH* component. We also measure the rest-frame ${\rm EW}_{\mathrm{[O\,III]}\lambda5007}$ from the best-fit spectral models described in Section~\ref{sec:specfit}. We find a strong anti-correlation between ${\rm EW}_{\mathrm{[O\,III]}\lambda5007}$ and the BH* fraction. At the same time, $L_{\rm [OIII]}$ is nearly independent of $\log L_{5100}$, suggesting that the \OIII{} emission primarily originates from the host-galaxy component rather than from the SMBH itself \citep{deGraaff_2025,PangYuxuan_2026}. This supports the decomposition of the LRD spectra into a host-galaxy component and a BH* component.

The third strongest correlation is between the Balmer-break strength and the BH* fraction, as already illustrated in the right panel of Figure~\ref{fig:BB_T}. In addition, the Balmer decrement shows a moderate correlation with the BH* fraction \citep{Chang_2026}. Finally, we note that $L_{\rm UV}$ is moderately correlated with both \Ha{} luminosity and \Ha{} equivalent width. This trend may be caused by two effects. First, it may partly reflect selection bias: because most LRDs are selected through their V-shaped SEDs, sources near the detection limit are preferentially identified when they have comparable rest-UV and rest-optical fluxes. Second, in our simplified decomposition, we assume that most of the rest-UV emission originates from the host galaxy and model the BH* component as a blackbody. However, as suggested by \citet{Sun_2026_BH*}, the BH* component may also contribute to the rest-UV emission.

Furthermore, we find an anti-correlation between stellar mass and BH* fraction. While this trend may be a trivial outcome from our spectral decomposition, i.e., larger BH* fraction corresponds to weaker host emission, it provides a cross-check on our mass estimates and spectral decomposition.

Overall, our toy decomposition model provides a simple but useful phenomenological description of the continuum diversity of LRDs within the BH* framework, in which the observed spectra can be interpreted as a combination of a blackbody-like BH* continuum and a host-galaxy component. Nevertheless, we have not tested, proven, or ruled out other possible scenarios. A more comprehensive investigation of the physical nature of LRDs is left as important future work.

\subsection{Non-detection for stellar absorption signals}

Within the BH* framework, the red optical continuum may arise from a dense blackbody atmosphere enshrouding the black hole, which could in principle produce absorption features similar to those seen in cool stellar atmospheres. This possibility has recently been examined by \citet{WangBingjie_2026_LRDwater}, who reported a prominent rest-frame $\sim1.4\,\micron$ water absorption feature in two $z\sim2$ LRDs among the coolest objects in their sample, with $T\lesssim3000$\,K. \citet{LinXiaojing_2026_localLRD} also detected extremely strong absorption lines in the local LRD ``The Egg'', which may indicate the presence of a cool, metal-enriched gas envelope.

We first examine whether stellar absorption features such as Mg\,I\,b, Na\,I\,D, and TiO can be identified in the LRD spectra. Although some broad continuum similarities are present, we do not detect these absorption features convincingly. This search is complicated by the presence of emission lines, as well as by the relatively low spectral resolution and large uncertainties of the spectra.

We then search for possible H$_2$O absorption in the LRD spectra. In the current NEXUS LRD sample, the only source that simultaneously satisfies the wavelength-coverage requirement and has sufficient SNR for this test is our lowest-redshift LRD, ID 93986, at $z=2.346$. Following \citet{WangBingjie_2026_LRDwater}, we construct a grid of dwarf templates from the PHOENIX stellar atmosphere models \citep{Allard_2012_Phoenix}, spanning effective temperature $2000<T<5500$\,K in steps of 100\,K, metallicity $-4.0<[{\rm M/H}]<0.5$ in steps of 0.5\,dex, and surface gravity $0<\log g<5.5$ in steps of 0.5\,dex. We model the LRD spectrum as a non-negative linear combination of two dwarf components, with both components allowed to vary freely within this template grid.

\begin{figure}
\epsscale{1.15}
\plotone{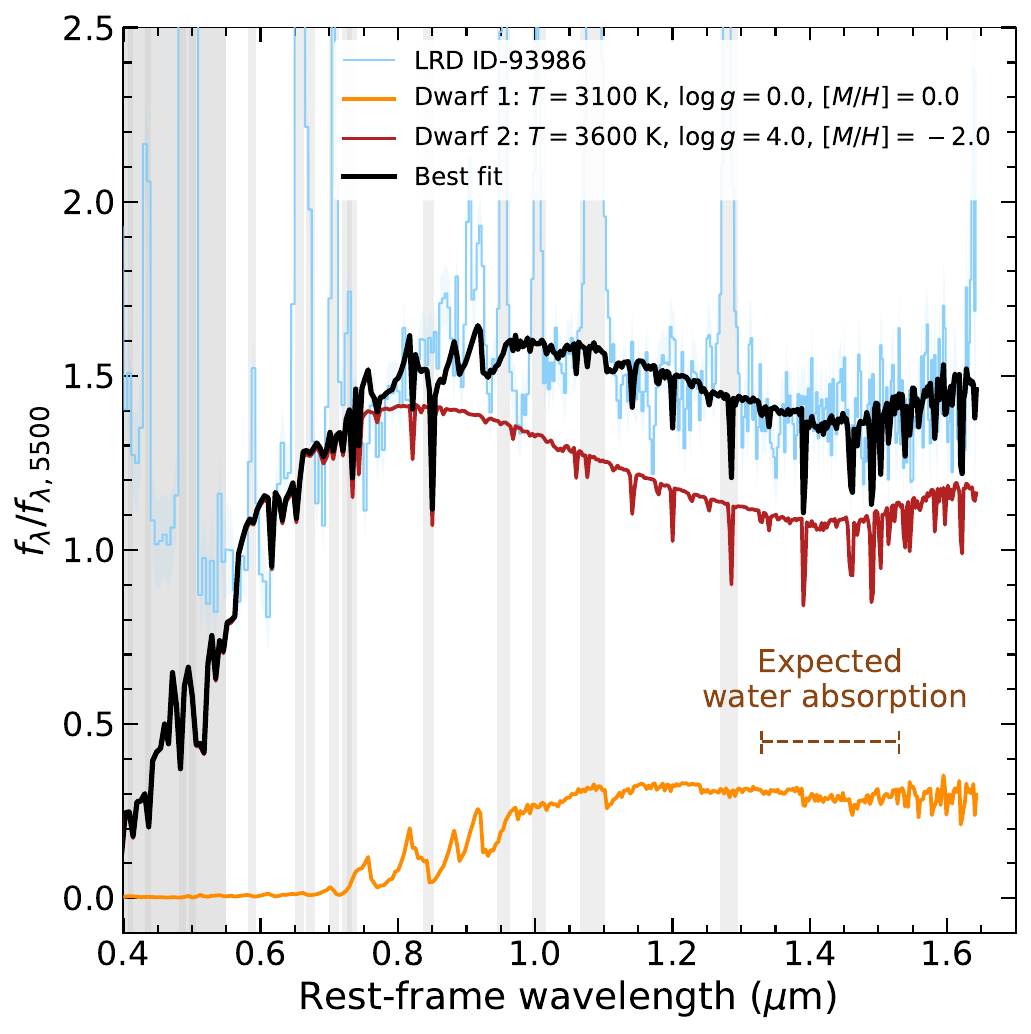}
\caption{
Search for possible water absorption in LRD ID 93986 at $z=2.346$. The rest-frame LRD spectrum is fitted with a two-component dwarf model based on PHOENIX stellar atmosphere templates \citep{Allard_2012_Phoenix}, where the template grid spans effective temperature $T$, metallicity $[{\rm M/H}]$, and surface gravity $\log g$. The gray shaded regions are excluded from the fit. The orange and red curves show the two fitted dwarf components, and the black curve shows their summed best-fit model. No significant water absorption signal is detected. We note that some relatively warm, high-$\log g$ dwarf templates can produce a trough-like spectral shape similar to the expected water absorption feature, introducing a degeneracy that makes a robust detection challenging.
}\label{fig:absorption}
\end{figure}

The best fit is shown in Figure~\ref{fig:absorption}. A two-dwarf model with temperatures of $T=3100$\,K and $T=3600$\,K can reproduce the observed spectrum well. Based on the fitting results, we note that the tentative trough around $\sim1.4\,\mu$m is explained by a relatively warm dwarf with high surface gravity, which produces a spectral shape similar to the expected water absorption feature. This introduces a degeneracy between a warm, high-$\log g$ dwarf and a cooler dwarf ($T<3000$\,K) with genuine water absorption. Because metallicity and surface gravity can strongly modify the spectral shape at fixed temperature, the current spectral quality and wavelength coverage do not provide evidence for a detectable water absorption feature.

Although we do not detect clear stellar-atmosphere-like absorption features in the NEXUS sample, this non-detection does not rule out the BH* scenario. Host-galaxy dilution, the limited spectral resolution and depth of the current data, and incomplete wavelength coverage can all weaken or obscure such features. The fact that a two-dwarf-component model can reproduce the observed spectrum may be qualitatively consistent with the possibility of multiple temperature layers in the BH* scenario. Future JWST observations optimized to target these absorption bands with higher SNR and more suitable wavelength coverage are essential for testing this aspect of the BH* model.


\section{Summary} \label{sec:sum}

In this work, we present a photometric and spectroscopic census of LRDs at $2.3<z<7.5$ using JWST/NIRCam imaging and NIRSpec/MSA PRISM spectroscopy from the ongoing NEXUS program. We apply three complementary photometric selection methods: V-shape color selection, V-shape slope selection, and Red-color selection. Together, these selections identify 463 photometric LRD candidates over the current NEXUS-Wide imaging footprint. From NIRSpec/MSA spectroscopy in the central Deep tier, we identify 36 spectroscopically confirmed LRDs. The combined photometric selection reaches a completeness of $\sim85\%$ based on the spectroscopically confirmed sample, while the purity is $\sim60\%$ for candidates with ${\rm F444W}<26$ mag. The main contaminants are strong emission-line galaxies and AGNs, as well as dwarf stars, while genuine LRDs not recovered by the photometric selections are mostly missed because of emission-line contamination in the broadband photometry. These results show that combining multiple photometric methods can yield a high-completeness LRD sample with quantifiable contamination.

We characterize the spectral properties of the spectroscopic LRD sample by modeling the \Hb+\OIII{} and \Ha{} regions in the PRISM spectra. Most LRDs, 31 out of 36, show resolved broad Balmer emission lines with ${\rm FWHM}>600~{\rm km~s^{-1}}$, and 90\% of these broad-line sources have ${\rm FWHM}>1000~{\rm km~s^{-1}}$. We estimate their host-galaxy stellar masses and discuss the large uncertainties associated with deriving black hole masses and bolometric luminosities from scaling relations calibrated for typical AGNs. We also provide the first constraint on the LRD abundance based on spectroscopic samples down to $\log L_{\rm H\alpha,broad}>42$ at $2.5<z<5$ in a regime where the spectroscopic completeness is well characterized. Our results suggest that low-luminosity LRDs at $z\sim2$--3 have a higher space density than currently inferred from ground-based searches targeting brighter systems. Furthermore, we present an exploratory clustering measurement of these spectroscopic LRDs. Although the uncertainty remains large, the inferred linear bias of $\sim6$ suggests that LRDs reside in dark matter halos of several $\times10^{11}\,h^{-1}M_\odot$.

The 36 spectroscopically confirmed LRDs include S-shape, V-shape, and more tentative L-shape sources, illustrating the broad spectral diversity of the population. They span a wide range of Balmer-break strengths, apparent break wavelengths, and emission-line properties. We find no clear redshift evolution in these main spectral properties over the redshift range probed by our sample. This diversity can be broadly explained by a red, blackbody-like rest-optical component combined with a mostly blue host-galaxy component, consistent with the BH* interpretation in which an accreting SMBH is enshrouded by dense gas. We also explore correlations among different spectral properties as empirical tests of this scenario. For example, the broad \Ha{} luminosity correlates with the rest-optical continuum luminosity, while the narrow \OIII{} emission shows a much weaker connection to the rest-optical continuum but a much stronger correlation with the inferred host-galaxy fraction.

Future NEXUS observations will improve these constraints. Additional NIRSpec/MSA spectroscopy in the Deep tier and NIRCam/WFSS spectroscopy in the Wide tier will increase the spectroscopic LRD sample size and improve measurements of completeness, luminosity functions, and clustering. The full NEXUS dataset will therefore provide a more complete and comprehensive view of the demographics, environments, and physical origin of LRDs.



\begin{acknowledgments}
Based on observations with the NASA/ESA/CSA James Webb Space Telescope obtained from the Barbara A. Mikulski Archive at the Space Telescope Science Institute, which is operated
by the Association of Universities for Research in Astronomy, Incorporated, under NASA contract NAS5-03127. Support for Program number JWST-GO-05105 was provided through a grant from the STScI under NASA contract NAS5-03127. 


\end{acknowledgments}

\vspace{5mm}
\facilities{JWST (NIRCam, NIRSpec)}

\software{
\texttt{Astropy} \citep{Astropy_2013,Astropy_2018,Astropy_2022}, 
\texttt{Matplotlib} \citep{Hunter_2007_Matplotlib}, 
\texttt{Numpy} \citep{Harris_2020_Numpy}, 
\texttt{scipy} \citep{SciPy_2020}
\texttt{photutils} \citep{Photoutils_2025}, 
\texttt{EAZY} \citep{Brammer_2008_EAZY}, 
          }


\bibliography{refs.bib}{}
\bibliographystyle{aasjournalv7}

\end{document}